\documentclass[jcappub,11pt]{article}
\pdfoutput=1 
\usepackage{jcappub} 
\usepackage[T1]{fontenc} 
\usepackage{url}
\usepackage{xcolor}
\usepackage{graphicx}
\usepackage{mathrsfs}
\usepackage[caption=false]{subfig}

\usepackage{tikz,xcolor,hyperref}

\definecolor{lime}{HTML}{A6CE39}
\DeclareRobustCommand{\orcidicon}{%
	\begin{tikzpicture}
	\draw[lime, fill=lime] (0,0) 
	circle [radius=0.16] 
	node[white] {{\fontfamily{qag}\selectfont \tiny ID}};
	\draw[white, fill=white] (-0.0625,0.095) 
	circle [radius=0.007];
	\end{tikzpicture}
	\hspace{-5mm}
}

\foreach \x in {A, ..., Z}{%
	\expandafter\xdef\csname orcid\x\endcsname{\noexpand\href{https://orcid.org/\csname orcidauthor\x\endcsname}{\noexpand\orcidicon}}
}

\title{\boldmath Probing massive neutrinos with the Minkowski functionals of the galaxy distribution}

\author[a,b]{Wei Liu\orcidA{},}
\author[a,b]{Aoxiang Jiang,}
\author[a,b,1]{and Wenjuan Fang \note{Corresponding author.}}

\affiliation[a]{CAS Key Laboratory for Research in Galaxies and Cosmology, Department of Astronomy, University of Science and Technology of China, Hefei, Anhui, 230026, P.R.China}
\affiliation[b]{School of Astronomy and Space Sciences, University of Science and Technology of China, Hefei, Anhui, 230026, P.R.China}

\emailAdd{lw980228@mail.ustc.edu.cn}
\emailAdd{jax9709@mail.ustc.edu.cn}
\emailAdd{wjfang@ustc.edu.cn}

\abstract{The characteristic signatures of massive neutrinos on large-scale structure (LSS), if fully captured, can be used to put a stringent constraint on their mass sum, $M_{\nu}$. Previous work utilizing N-body simulations has shown the Minkowski functionals (MFs) of LSS can reveal the imprints of massive neutrinos on LSS, provide important complementary information to two-point statistics, and significantly improve constraints on $M_{\nu}$. In this work, we take a step forward and apply the statistics to the biased tracers of LSS, i.e. the galaxies, and in redshift space. We perform a Fisher matrix analysis and quantify the constraining power of the MFs by using the Molino mock galaxy catalogs, which are constructed based on the halo occupation distribution (HOD) framework with parameters for the SDSS $M_r < -21.5$ and -22 galaxy samples. We find the MFs give tighter constraints on all of the cosmological parameters that we consider than the power spectrum. The constraints on $\Omega_{\mathrm{m}}, \Omega_{\mathrm{b}}, h, n_s, \sigma_8$, and $M_\nu$ from the MFs are better by a factor of 1.9, 2.9, 3.7, 4.2, 2.5, and 5.7, respectively, after marginalizing over the HOD parameters. Specifically, for $M_{\nu}$, we obtain a 1$\sigma$ constraint of 0.059 eV with the MFs alone for a volume of only $\left(1 h^{-1} \mathrm{Gpc}\right)^3$.
}

\begin{document}
\maketitle
\flushbottom

\section{Introduction}
Neutrino oscillation experiments have firmly established the existence of neutrino masses \cite{PhysRevLett.81.1562,PhysRevLett.89.011301,PhysRevLett.94.081801,PhysRevLett.101.131802}, which is the direct laboratory evidence for physics beyond the Standard Model. However, oscillation experiments are insensitive to the absolute scale of neutrino masses, thus to the neutrino mass sum $M_{\nu}$ \cite{GONZALEZGARCIA2016199}. Laboratory experiments like beta-decay and neutrinoless double-beta decay ($0\nu2\beta$) of nuclei can be used to measure the absolute neutrino mass scale \cite{gerbino2018status}. The beta-decay is robust and model-independent but less sensitive than $0\nu2\beta$ decay; while
 the $0\nu2\beta$ decay poses more stringent constraints on the neutrino mass scale at the price of model dependence \cite{2022arXiv220307377A}. Using the beta decay of tritium, the KATRIN experiment has measured an incoherent sum of mass states \cite{PhysRevLett.123.221802,2021arXiv210508533A}, which approximately translates to $M_{\nu}\lesssim 2.4 eV$ and is the best current experimental upper bound on the neutrino masses \cite{2022arXiv221112457A}.
 
 %Although upcoming laboratory experiments are expected to improve the sensitivity by an order of magnitude \cite{2001hep.ex....9033K}, determining the absolute mass scale is still a challenging task for laboratory experiments \cite{drexlin+13, Salas+18}.

 %The measured mass splittings, $\Delta m_{21}^{2} \equiv m_{2}^{2}-m_{1}^{2} \approx 7.49_{-0.17}^{+0.19} \times 10^{-5} \mathrm{eV}^{2}$ and  $\left|\Delta m_{31}^{2}\right| \equiv\left|m_{3}^{2}-m_{1}^{2}\right| \approx$ $2.484_{-0.048}^{+0.045} \times 10^{-3} \mathrm{eV}^{2}(1 \sigma)$ \cite{hierarchy},  imply that at least two of the mass eigenvalues are nonzero. Depending on the sign of $\Delta m_{31}^{2}$, there are two possible rankings for the three neutrino masses. Assuming the lightest mass to be zero, the minimum summed mass $M_{\nu}$ for the two rankings are approximately 0.06 eV (normal hierarchy with $m_1<m_2<m_3$) and 0.1eV (inverted hierarchy with $m_2>m_1>m_3$) \cite{Hannestad_2016}. However, determining the mass ranking and the absolute mass scale is generally challenging for laboratory experiments \cite{drexlin+13, Salas+18}.

Through a completely independent approach, cosmology offers a promising probe of the neutrino masses \cite{Hu+98, Abazajian+15, Dvorkin+19,2022JHEAp..36....1T}. The Big Bang theory predicts the existence of a cosmic neutrino background. With nonzero masses, cosmic neutrinos can change both the Universe's expansion and the evolution of its perturbations \cite{2006PhR...429..307L,2011ARNPS..61...69W}. In early Universe, they are relativistic and act as radiation. As the Universe cools down, they become non-relativistic and behave like matter. Keeping the total matter density today fixed, non-zero neutrino masses reduce the mass fraction of baryons and cold dark matter, which leads to a later epoch of matter-radiation equality when neutrinos are typically still relativistic. In addition, the large thermal velocities of neutrinos allow them to stream out of the cold dark matter potential wells, so they do not contribute to matter clustering, and the growth of structure is suppressed on scales smaller than their free-streaming scales. As a result, massive neutrinos leave observable imprints on both the cosmic microwave background (CMB) and large-scale structure (LSS).

The current tightest constraint on $M_{\nu}$ comes from cosmology: $M_{\nu}\lesssim 0.12$ eV at 95\% confidence level, obtained mainly by combining observations of the CMB anisotropies and the Baryonic Acoustic Oscillations (BAO) \cite{2020A&A...641A...6P,2021PhRvD.103h3533A}. Since the suppression of growth of structure by massive neutrinos is strongest on small scales and at low redshifts \cite{Brinckmann_2019,2018JCAP...03..049L}, tighter constraints are expected by including more low-redshift LSS data on small scales. Current and upcoming galaxy surveys such as DESI \footnote{\href{http://www.desi.lbl.gov}{http://www.desi.lbl.gov}}, PFS \footnote{\href{http://pfs.ipmu.jp}{http://pfs.ipmu.jp}}, Roman Space Telescope \footnote{\href{http://wfirst.gsfc.nasa.gov}{http://wfirst.gsfc.nasa.gov}}, Euclid \footnote{\href{http://sci.esa.int/euclid}{http://sci.esa.int/euclid}} and CSST \cite{CSST,2019ApJ...883..203G}\footnote{\href{http://nao.cas.cn/csst}{http://nao.cas.cn/csst}}, will provide high-precision measurements of the 3D clustering of galaxies. Besides, simulations have made considerable progress in modeling nonlinear structure formation \cite{Brandbyge_2008,Banerjee&Dalal16,Villaescusa_Navarro_2013,Villaescusa_Navarro_2018,Castorina_2015,PhysRevD.90.045022,Castorina_2014,10.1111/j.1365-2966.2011.19488.x,2016PhRvD..93f3515U} in cosmologies with massive neutrinos. With the development in both observations and simulations, it is promising to unlock the information beyond linear scales and tightly constrain $M_{\nu}$.

However, parameter degeneracies exist in the most commonly used statistics of LSS, the two-point correlation function or its Fourier transform, the power spectrum, and limit their constraining power on $M_{\nu}$. Particularly, the $\sigma_8$ and $M_{\nu}$ degeneracy is reported to be one of the most prominent \cite{Villaescusa_Navarro_2018}. 
In addition, a large amount of information leaks from the two-point statistics due to the non-Gaussianities induced by gravity.  Statistical tools capable of uncovering the important missed information are needed to break the parameter degeneracies and obtain tighter constraints on $M_{\nu}$. Multiple efforts have been made in this direction. The bispectrum is demonstrated to help break the degeneracy between $M_{\nu}$ and $\sigma_{8}$ \cite{2020JCAP...03..040H} and tighten the constraints on $M_{\nu}$ \cite{Chudaykin_2019}. Void statistics have been found to be capable of capturing some characteristic effects of massive neutrinos on LSS \cite{Kreisch_2019,Massara_2015}. In order to extract the information embedded in low-density regions like cosmic voids, marked power spectrum that emphasizes low-density regions has been proposed to place tight constraints on $M_{\nu}$ \cite{2021PhRvL.126a1301M}. The one-point probability distribution function (PDF) of the matter density field has been found to be highly complementary to the matter power spectrum on mildly non-linear scales, and help tighten the constraints on $M_{\nu}$ \cite{2020MNRAS.495.4006U}. 

In our previous work \cite{2022JCAP...07..045L}, we explored in a quantitive way the potential of using LSS's morphological properties to constrain $M_{\nu}$, which are fully characterized by 4 Minkowski functionals (MFs). 
According to Hadwidger's theorem \cite{Hadwiger_1957}, for a pattern in n-dimensional \textcolor{black}{space}, its morphological properties \textcolor{black}{(defined as those satisfying motional-invariance and additivity)} can be fully described by (n+1) Minkowski functionals (MFs) \footnote{\textcolor{black}{The more rigorous and mathematical description of Hadwiger's theorem is given in \cite{1994A&A...288..697M} as: any additive, motion invariant and conditionally continuous functional $\mathscr{F}$ on a body $A$ in $d$ dimension is a linear combination of the $d+1$ Minkowski functional $\mathscr{F}(A)=\sum_{\i=0}^d c_i V_i(A)$, with real coefficients $c_i$ independent of $A$.}}. In 3D, the 4 MFs are, respectively, the pattern's volume, surface area, integrated mean curvature, and Euler characteristic (or genus). The MFs can principally probe all orders of statistics \cite{1994A&A...288..697M,10.1046/j.1365-8711.1999.02912.x}, thus are expected to provide supplementary information for the power spectrum. For a more detailed review of the MFs and their application in cosmology, please refer to \cite{2022JCAP...07..045L,2021MNRAS.tmp.2528L,2020arXiv201104954M,2020arXiv201200203M} and references therein. Using the Quijote simulation \cite{2020ApJS..250....2V}, we quantified the constraints on cosmological parameters from the MFs. Depending on whether massive neutrinos are included in the density field (the `m' field) or not (the `cb' field), we found the constraint on $M_{\nu}$ from the MFs was 48 or 4 times better than that from the power spectrum. When the MFs are combined with the power spectrum, they can improve the constraint on $M_{\nu}$ from the latter by a factor of 63 for the `m' field and 5 for the `cb' field. Although constraints from the MFs of the `cb' field are weaker than those of the `m' field, they have a closer connection with the MFs of the observed galaxies, which are the focus of this work.

In this paper, we forecast the constraints on $M_{\nu}$ jointly with other cosmological parameters from the MFs of the redshift-space galaxy distribution for the first time. Using the Molino mock galaxy catalogs \cite{2021JCAP...04..029H}, we obtain accurate simulation-based modeling of the MFs down to mildly nonlinear scales for the redshift-space galaxy distribution and reliable Fisher matrix forecasts with a large number of samples. We also make a direct comparison with other statistics that have been applied to these mock catalogs, such as the power spectrum, the bispectrum monopole \cite{2021JCAP...04..029H}, and the marked power spectrum \cite{2022arXiv220601709M}. Notably, we detailedly compare the MFs and the power spectrum on parameter degeneracies present in each statistic and parameter constraints from each statistic. Constraints from their combination are also presented.

This paper is organized as follows. Section \ref{models} describes the Quijote simulation suite and the Molino mock galaxy catalogs. We then describe the measurement of Minkowski functionals in Section~\ref{sec:mfs}. The effects of redshift space distortion, shot noise, and galaxy bias are briefly discussed in Section~\ref{sec:RSD_Shot_Bias}. The Fisher information matrix formalism used to calculate parameter constraints is explained in Section~\ref{sec:fisher}, and the constraints on cosmological parameters from the MFs, the power spectrum, and their combination are given in Section \ref{sec:result}. Section~\ref{sec:discuss} discusses the results obtained and makes a comparison with other works based on the Molino catalogs and our previous work. Finally, we conclude in Section \ref{conclusions} and discuss some subtleties of our forecast in the appendices. 

\section{The Quijote simulations and Molino mock galaxy catalogs}
\label{models}

In this work, we use the Molino mock galaxy catalogs \cite{2021JCAP...04..029H} to quantify the information content embedded in the MFs of galaxy distributions with the Fisher matrix formalism, these mock galaxy catalogs are constructed based on the Quijote\footnote{https://github.com/franciscovillaescusa/Quijote-simulations} simulation suite \cite{2020ApJS..250....2V}. The Quijote suite is a set of over 44,000 N-body simulations spanning over 7,000 cosmological models. It is designed to quantify cosmological observables' information content and train machine learning algorithms. The Quijote simulation suite is run using the TreePM+SPH code GADGET-III \cite{10.1111/j.1365-2966.2005.09655.x} with a cosmological volume of $1\left(h^{-1} \mathrm{Gpc}\right)^{3}$ and $512^3$ CDM particles (plus $512^3$ neutrino particles for cosmologies with massive neutrinos), where cosmic neutrinos are modeled using the traditional particle-based method \cite{Viel_2010,Brandbyge_2008}, and degenerate masses of massive neutrinos are assumed. For cosmologies with massive neutrinos and their fiducial counterparts with massless neutrinos, the initial conditions (ICs) are generated at $z = 127$ using the rescaling method \cite{10.1093/mnras/stw3340} employing the Zel'dovich approximation. For all other cosmologies used in this work, the initial conditions are generated using second-order perturbation theory (2LPT) instead. 

The fiducial model has the cosmological parameter values set to be in good agreement with the latest Planck constraints \cite{2020A&A...641A...6P}: the matter density parameter $\Omega_{\mathrm{m}}=0.3175$, the baryon density parameter $\Omega_{\mathrm{b}}=0.049$, the dimensionless Hubble constant $h=0.6711$, the spectral index $n_{s}=0.9624$, the root-mean-square amplitude of the linear matter fluctuations at $8h^{-1} \rm{Mpc}$ $\sigma_{8}=0.834$, the sum of neutrino masses $M_{\nu}=0.0$ eV, and the dark energy state parameter $w=-1$. For the fiducial model, 15000 realizations are run for the accurate estimate of covariance matrices, while for the models where only one of the parameters $\Omega_m,\Omega_b,h,n_s,\sigma_8$ varies at a time, the three models with massive neutrinos, and the fiducial model with Zel'dovich ICs, 500 realizations are run to precisely estimate the derivatives along cosmological parameters. Specifications of the simulations associated with mock galaxy catalogs used in this work can be found in Table \ref{tab:s}.

The Molino mock galaxy catalogs are built upon the halo catalogs of the Quijote simulations at $z=0$, which are generated on the CDM+baryon distribution using a Friends-of-Friends (FoF) algorithm \cite{1985ApJ...292..371D} with linking length parameter $b = 0.2$. To populate these halos with galaxies, the standard HOD model from \cite{Zheng_2007} is used. In the HOD framework, the probability of a given halo that hosts $N_{\text{gal}}$ galaxies depends only on its halo mass $M_{h}$. More specifically, the mean number of galaxies in a halo is the sum of the mean number of central and satellite galaxies,
\begin{equation}
\left\langle N_{\text {gal }}\right\rangle=\left\langle N_{\mathrm{cen}}\right\rangle+\left\langle N_{\mathrm{sat}}\right\rangle,
\end{equation}
the mean central galaxy occupation is given by
\begin{equation}
\label{eq:central_hod}
\left\langle N_{\text{cen}}\right\rangle=\frac{1}{2}\left[1+\operatorname{erf}\left(\frac{\log M_h-\log M_{\min }}{\sigma_{\log M}}\right)\right],
\end{equation}
where $M_{\text{min}}$ is the minimum mass of the host halos of central galaxies and $\sigma_{log M}$ is the width of the transition of $\left\langle N_{\text{cen}}\right\rangle$ from 0 to 1.
The mean satellite galaxy occupation is specified by 
\begin{equation}
\label{eq:satellite_hod}
\left\langle N_{\mathrm{sat}}\right\rangle=\left\langle N_{\mathrm{cen}}\right\rangle\left(\frac{M_h-M_0}{M_1}\right)^\alpha ,
\end{equation}
 it follows a power law with slope $\alpha$, and is also controlled by the halo mass cut-off $M_0$ for satellite galaxy occupation and the mass scale $M_1$ for quantifying the difference between $M_h$ and $M_0$. The fiducial values of the HOD parameters for the Molino catalogs are set as
\begin{equation}
\left\{\log M_{\min }, \sigma_{\log M}, \log M_0, \alpha, \log M_1\right\}=\{13.65,0.2,14.0,1.1,14.0\},
\end{equation}
which are not exactly equal to the best-fit HOD parameters for the SDSS $M_r<-21.5$ and -22 samples from \cite{Zheng_2007}, because of the halo mass limit of the Quijote simulations \footnote{For more details about the fiducial values of the HOD parameters, see Section~3 of \cite{2021JCAP...04..029H}}. There are also five pairs of non-fiducial HOD models (a pair per HOD parameter); for each pair, only one HOD parameter varies above and below the fiducial value by the step sizes
\begin{equation}
\left\{\Delta \log M_{\min }, \Delta \sigma_{\log M}, \Delta \log M_0, \Delta \alpha, \Delta \log M_1\right\}=\{0.05,0.02,0.2,0.2,0.2\}.
\end{equation}
These step sizes were chosen to make the derivatives for the power spectrum and the bispectrum converged \cite{2021JCAP...04..029H}. And they have also been tested to give convergent derivatives for the marked power spectrum \cite{2022arXiv220601709M}. Hence we anticipate the derivatives of the MFs w.r.t. HOD parameters estimated with these models are also convergent.

\begin{center}
	\small
	\begin{table}[tbp]
	\begin{tabular}{|cccccccccc|}
		\hline Name & $M_{\nu}$ & $\Omega_{m}$ & $\Omega_{b}$ & $h$ & $n_{s}$ & $\sigma_{8}$ & $w$ & ICs & realizations \\
		\hline Fiducial & 0.0 & 0.3175 & 0.049 & 0.6711 & 0.9624 & 0.834 & -1 & 2LPT & 15000 \\
		Fiducial ZA & 0.0 & 0.3175 & 0.049 & 0.6711 & 0.9624 & 0.834 & -1 & Zel'dovich & 500 \\
		$M_{\nu}^{+}$ & \underline{0.1} & 0.3175 & 0.049 & 0.6711 & 0.9624 & 0.834 & -1 & Zel'dovich & 500 \\
		$M_{\nu}^{++}$ & \underline{0.2} & 0.3175 & 0.049 & 0.6711 & 0.9624 & 0.834 & -1 & Zel'dovich & 500 \\
		$M_{\nu}^{+++}$ & \underline{0.4} & 0.3175 & 0.049 & 0.6711 & 0.9624 & 0.834 & -1 & Zel'dovich & 500 \\
		$\Omega_{m}^{+}$ & 0.0 & \underline{0.3275} & 0.049 & 0.6711 & 0.9624 & 0.834 & -1 & 2LPT & 500 \\ 
		$\Omega_{m}^{-}$ & 0.0 & \underline{0.3075} & 0.049 & 0.6711 & 0.9624 & 0.834 & -1 & 2LPT & 500 \\ 
		$\Omega_{b}^{++}$ & 0.0 & 0.3175 & \underline{0.051} & 0.6711 & 0.9624 & 0.834 & -1 & 2LPT & 500 \\ 
		$\Omega_{b}^{--}$ & 0.0 & 0.3175 & \underline{0.047} & 0.6711 & 0.9624 & 0.834 & -1 & 2LPT & 500 \\ 
		$h^{+}$ & 0.0 & 0.3175 & 0.049 & \underline{0.6911} & 0.9624 & 0.834 & -1 & 2LPT & 500 \\ 
		$h^{-}$ & 0.0 & 0.3175 & 0.049 & \underline{0.6511} & 0.9624 & 0.834 & -1 & 2LPT & 500 \\ 
		$n_{s}^{+}$ & 0.0 & 0.3175 & 0.049 & 0.6711 & \underline{0.9824} & 0.834 & -1 & 2LPT & 500 \\ 
		$n_{s}^{-}$ & 0.0 & 0.3175 & 0.049 & 0.6711 & \underline{0.9424} & 0.834 & -1 & 2LPT & 500 \\ 
		$\sigma_{8}^{+}$ & 0.0 & 0.3175 & 0.049 & 0.6711 & 0.9624 &  \underline{0.849} & -1 & 2LPT & 500 \\ 
		$\sigma_{8}^{-}$& 0.0 & 0.3175 & 0.049 & 0.6711 & 0.9624 &  \underline{0.819} & -1 & 2LPT & 500 \\
		\hline
		
	\end{tabular}
    \caption{\label{tab:s} The subsets of the Quijote simulation suites used in this work. 15000 fiducial simulations of the Quijote suite are used for estimating the covariance matrices, and 500 simulations each for 14 different cosmologies are used for calculating derivatives of observables with respect to cosmological parameters.}
    \end{table}
\end{center}

For the estimate of the derivatives w.r.t. cosmological parameters, 5 sets of HOD realizations with different random seeds are generated applying the fiducial HOD model, while for the estimate of the derivatives w.r.t. HOD parameters, 5 sets of HOD realizations are generated using the corresponding non-fiducial HOD models from 500 fiducial simulations. Redshift-space distortion (RSD) is applied along three different axes (the x, y, and z-axis) for these HOD catalogs. For mocks used to estimate the covariance matrix, one set of HOD realization is generated for each of the 15000 simulations at the fiducial cosmology, and RSD is applied along the z-axis. To summarize, a total of 180000 samples (7500 samples per non-fiducial model) are used to estimate the derivatives w.r.t. cosmological and HOD parameters, and 15000 samples are generated for estimating the covariance matrix.

\section{Measurement of the Minkowski functionals}
\label{sec:mfs}

For a spatial pattern in three-dimensional space, we can study its morphology using four Minkowski functionals, namely its volume ($V_{0}$), surface area ($V_{1}$), integrated mean curvature ($V_{2}$), and Euler characteristic ($V_{3}$). Instead of the functionals themselves, their spatial densities are more commonly used for convenient comparison between samples with different volumes. Therefore, in this work, we divide the four Minkowski functionals with the volume of the simulation box. Thus $V_{0}$ refers to the volume fraction, while $V_{1}$, $V_{2}$ and $V_{3}$ are the surface area, the integrated mean curvature, and the Euler characteristic per unit volume, respectively. To measure the MFs, we first interpolate the positions of mock galaxies onto a grid with $N_{grid}=360$ using the piecewise cubic spline (`PCS') \footnote{We use the routine provided by Pylians (Python libraries for the analysis of numerical simulations): https://pylians3.readthedocs.io/en/master/, other mass assignment schemes are also available, such as `NGP' (nearest grid point), `CIC' (cloud-in-cell) and `TSC' (triangular-shape cloud)} mass assignment scheme. We have tested other choices of $N_{grid}$ and mass assignment schemes, and find that $N_{grid}=360$ and `PCS' already give a convergent result. \textcolor{black}{The galaxy number density field \footnote{We denote the number density field interchangeably by $\rho(\boldsymbol{x})$ or $n(\boldsymbol{x})$, the subscript $g$ is added to differentiate the galaxy field and particle field.} is then transformed to the galaxy density contrast field with $\delta (\boldsymbol{x})=(\rho(\boldsymbol{x})-\bar{\rho})/\bar{\rho}$. Hereafter, the word ``density'' will refer to density contrast most of the time.} Since the pixel size, defined as $L_{box}/N_{grid}$, is about $2.8 h^{-1} \rm{Mpc}$ in this work, which is smaller than the adopted smoothing scale $R_G$ by at least a factor of seven. It is anticipated that the effect of finite pixel size \cite{2014ApJS..212...22K} can be neglected for the galaxy field interpolated onto this grid \cite{2005ApJ...633....1P}.

To suppress the shot noise, \textcolor{black}{the galaxy density contrast fields} are smoothed with a Gaussian window function of radius $R_G \geq 20 h^{-1} \rm{Mpc}$. For the Gaussian window function, the volume of the window is related to the smoothing radius $R_G$ as \textcolor{black}{$V(R_G)=(4\pi)^{3/2}R^3_G$} (see Eq.~\ref{eq:sigma_0_shot}, while for a top-hat window function, $V(R)=4\pi R^3/3$ \cite{mo_van_2010}). Because the galaxy number density of the fiducial HOD model is $\bar{n}_g \sim 1.63 \times 10^{-4} h^3 / \mathrm{Mpc}^3$, smoothing scales satisfying $R_G \geq 20 h^{-1} \rm{Mpc}$ make sure that at least $\sim 60$ galaxies are covered by every Gaussian window. The mean galaxy separation $\bar{d}$ is about $18h^{-1}\rm{Mpc}$ for the Molino catalogs. In analyses of observational data using the MFs or the genus alone\footnote{the genus $g$ is related with the Euler characteristic $V_3$ by $g=1-V_3$ \cite{1997ApJ...482L...1S}}, the choice of $R_G=\bar{d}/\sqrt{2}$ \cite{1994ApJ...420..525V,1989ApJ...340..625G} or $R_G=\bar{d}$ \cite{hikage2003minkowski,2005ApJ...633....1P} is frequently used. \textcolor{black}{In \cite{2014ApJS..212...22K}, the smoothing scale $R_G=\bar{d}$ is recommended for the genus topology analysis, and they found the shot noise effects diminish greatly when biased objects are used. We will discuss the effect of shot noise on the MFs more detailedly in Section~\ref{sec:shot_noise_correction}}. On the other hand, the corresponding wavenumber for the scale $R_G$ is approximately $2\pi/R_G$, which is smaller than $0.5 h \rm{Mpc}^{-1}$ when $R_G\geq 20 h^{-1} \rm{Mpc}$. This is larger than the scales where the N-body simulations are not converged \cite{2020ApJS..250....2V}.

To extract more information and break parameter degeneracies, \textcolor{black}{the density contrast fields} are smoothed with four different smoothing scales $R_G=20,\ 25,\ 30,\ 35h^{-1}\rm{Mpc}$, whose MFs are combined for parameter constraints. Although stronger constraints can be obtained when more smoothing scales are included, we need to keep the length of the data vector not so long that the estimate of the covariance matrix may be inaccurate. In addition, as will be explained in section~\ref{sec:fisher} and Appendix~\ref{sec:non-Gaussian}, we choose the scales $R_G < 40 h^{-1} \rm{Mpc}$ to avoid introducing intractable non-Gaussianities to the likelihood of the data vector. Then we measure the MFs for the excursion sets of the smoothed density field, which are regions with density above a given density threshold \footnote{For helpful visualization and detailed description of these excursion sets with different density thresholds, see \cite{2022JCAP...07..045L}}. Two complementary formulae suitable for numerically calculating the MFs through differential geometry (Koenderink invariant) and integral geometry (Crofton's formula) are derived in \cite{1997ApJ...482L...1S}. We have checked that the MFs calculated with different formulae give consistent results. Hereafter, we only present the results obtained using Crofton's formula.

\begin{figure}[tbp]
	\centering 
	\includegraphics[width=1.0\textwidth]{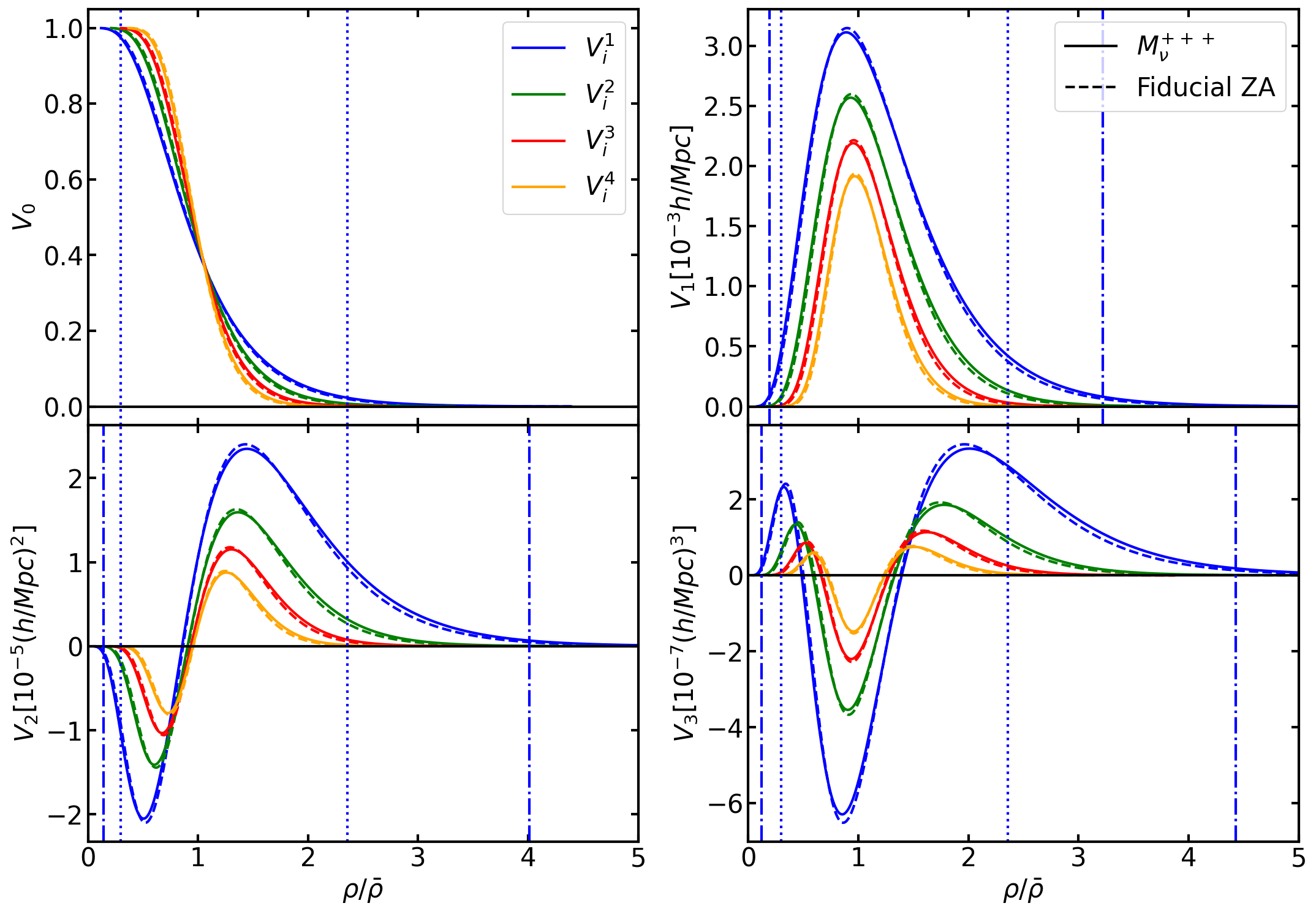}
	\caption{\label{fig:Vis} The MFs for the $M_{\nu}^{+++}$ (solid lines) and Fiducial ZA (dashed lines) cosmologies are shown as functions of density threshold $\rho / \overline{\rho}$. The MFs with the four different smoothing scales $R_G=20,\ 25,\ 30,\ 35h^{-1}\rm{Mpc}$ are labeled as $V_i^{1,2,3,4}$, and plotted in blue, green, red, and orange, respectively. To manifest our motivation for the new threshold binning scheme discussed in Section~\ref{sec:fisher_deri}, we show the difference in the two schemes for $R_G=20 h^{-1}\rm{Mpc}$ as an example. The two blue dotted vertical lines in each panel indicate the region that will be used if we use the old threshold binning scheme. In comparison, the two blue dash-dotted vertical lines in the panel for $V_1$, $V_2$, and $V_3$ indicate the region used in this forecast with our new threshold binning scheme. This new scheme is adaptive for each of the four MFs, considering the varied threshold range where different $V_i$s have a notable signal. }
\end{figure}

In figure~\ref{fig:Vis}, the four MFs with four different smoothing scales $R_G=20,\ 25,\ 30,\ 35h^{-1}\rm{Mpc}$ are plotted for the $M_{\nu}^{+++}$ (solid lines) and Fiducial ZA (dashed lines) cosmologies. The MFs for other models have similar shapes as plotted in this figure, and how they change when the smoothing scale is increased is also alike. Thus, we only show the curves for the $M_{\nu}^{+++}$ and Fiducial ZA cosmologies here. We refer to \cite{2022JCAP...07..045L} for a detailed interpretation of the four MFs and how they change when the sum of neutrino masses $M_{\nu}$ varies. Here we will focus on explaining the change of the four MFs with the increase of $R_G$.

When we smooth the density contrast field with larger scales, the variance of the field becomes smaller, and the distribution of the density field values concentrates more around the mean $\bar{\rho}$. Therefore, the curves of the four MFs shift towards $\rho/\bar{\rho}\sim 1$ from both the low and high threshold sides. At the same time, the amplitudes of $V_i$s drop except for $V_0$. $V_0$ is the volume fraction of the excursion set with the threshold $\rho/\bar{\rho}$, it transits from 1 to 0 with the increase of $\rho/\bar{\rho}$. Raising the smoothing scale only reduces the width of this transition. In contrast, when using larger $R_G$, the isodensity contours have fewer details and become smoother. Hence the area of the contours decreases and the amplitude of $V_1$ drops. $V_2$ and $V_3$ are proportional to the surface integral of the local Minkowski functionals 
\begin{equation}
\begin{aligned}
& v_2^{(\mathrm{loc})}(\rho/\bar{\rho}, \boldsymbol{x})=\frac{1}{6 \pi}\left(\frac{1}{R_1}+\frac{1}{R_2}\right) \\
& v_3^{(\mathrm{loc})}(\rho/\bar{\rho}, \boldsymbol{x})=\frac{1}{4 \pi} \frac{1}{R_1 R_2}
\end{aligned}
\end{equation}
where $R_1$ and $R_2$ are the two principal radii of curvature for the isodensity contour with the density threshold $\rho/\bar{\rho}$ at the position $\boldsymbol{x}$. For larger smoothing scales, we find $|R_1|$ and $|R_2|$ are larger on average. Thus, the amplitudes of $V_2$ and $V_3$ drop due to both the reduced surface area and the enlarged principal radii of curvature.

%How the derivative change w.r.t. $R_G$ can reflect its scale dependence, which will help us understand the scale dependence of parameter constraints.

\section{Effects of redshift space distortion, shot noise, and galaxy bias on the MFs}
\label{sec:RSD_Shot_Bias}

Our previous work \cite{2022JCAP...07..045L} was based on cold dark matter (plus massive neutrino) particle distribution in real space, while in this work we focus on redshift space galaxy distribution. Redshift space distortion (RSD), shot noise, and galaxy bias all play a role in producing the difference between the MFs of the two different distributions. We will briefly discuss them one by one.

Since we have used an untraditional density threshold $\rho/\bar{\rho}$ in this work, it will be helpful to first clarify its difference with the more frequently used (especially in papers working on deriving analytical expressions for the MFs) threshold parameter $\nu$ and $\nu_A$. By definition, $\rho/\bar{\rho}\equiv\delta +1$, $\nu \equiv \delta/\sigma$ ($\sigma$ is the variance of the smoothed density contrast field), and $\nu_A$ is defined to make sure the excursion set has the same volume fraction as a corresponding Gaussian field ($f_A=\frac{1}{\sqrt{2\pi}}\int_{\nu_A}^{\infty}e^{-t^2/2}dt$, $f_A$ is the fractional volume of the field above $\nu_A$).  Therefore, any factor, physical or unphysical, that changes the variance of the density field will rescale the threshold $\rho/\bar{\rho}$, and the curves of the MFs as a function of $\rho/\bar{\rho}$ will be shifted, expanded or compressed. That is, more information is embedded in the MFs as a function of $\rho/\bar{\rho}$ but systematics also have a larger impact on $V_i(\rho/\bar{\rho})$. For $V_i(\nu)$, the variation of $\sigma$ won't rescale the threshold since the threshold is already normalized by $\sigma$. The threshold parameter $\nu_A$ is more interesting because $V_i(\nu_A)$ is invariant under a local monotonic transformation of the density field \cite{2013MNRAS.435..531C}, and the biasing scheme $\delta_g=b\delta+\frac{b_2}{2}(\delta^2-\langle\delta^2\rangle)$ won't introduce non-Gaussian corrections to the MFs \cite{2003ApJ...584....1M}. We will comprehensively compare the three different threshold parameters from aspects of physical information content and robustness to systematics in another ongoing project of ours.

\subsection{Redshift space distortion}

The effect of redshift space distortion on statistics of isodensity contour, including the Minkowski functionals was first studied in \cite{1996ApJ...457...13M}, where they found these statistics in redshift space as a function of $\nu_A$ have the same shape as in real space, the redshift space distortion only affects amplitudes of these statistics in the Gaussian limit. The non-Gaussian RSD effect was detailedly studied in \cite{2013MNRAS.435..531C},  but we will only focus on the dominant Gaussian RSD effect here and refer the interested readers to their work for more details. Take the genus statistics as an example, the expression of genus $G^{(s)}$ in redshift space is given by 
\begin{equation}
G^{(s)}(\nu_A)=\frac{3\sqrt{3}}{2}\sqrt{\frac{C_1}{C_0}}(1-\frac{C_1}{C_0})G^{(r)}(\nu_A), 
\end{equation}
where
\begin{equation}
\frac{C_1}{C_0}=\frac{1}{3}\frac{1+(6/5)fb^{-1}+(3/7)(fb^{-1})^2}{1+(2/3)fb^{-1}+(1/5)(fb^{-1})^2},
\end{equation}
$f$ is the growth rate of the structure, $b$ is the linear bias parameter, and $G^{(r)}(\nu_A)$ is the genus in real space. With these equations, we can see the sensitivity of the MFs to $fb^{-1}$.  In a recent work \cite{2023arXiv230803086J}, a new statistic with a very close relation with Minkowski functionals was derived for stronger constraining power on $fb^{-1}$.

\begin{figure}[tbp]
	\centering 
	\includegraphics[width=1.0\textwidth]{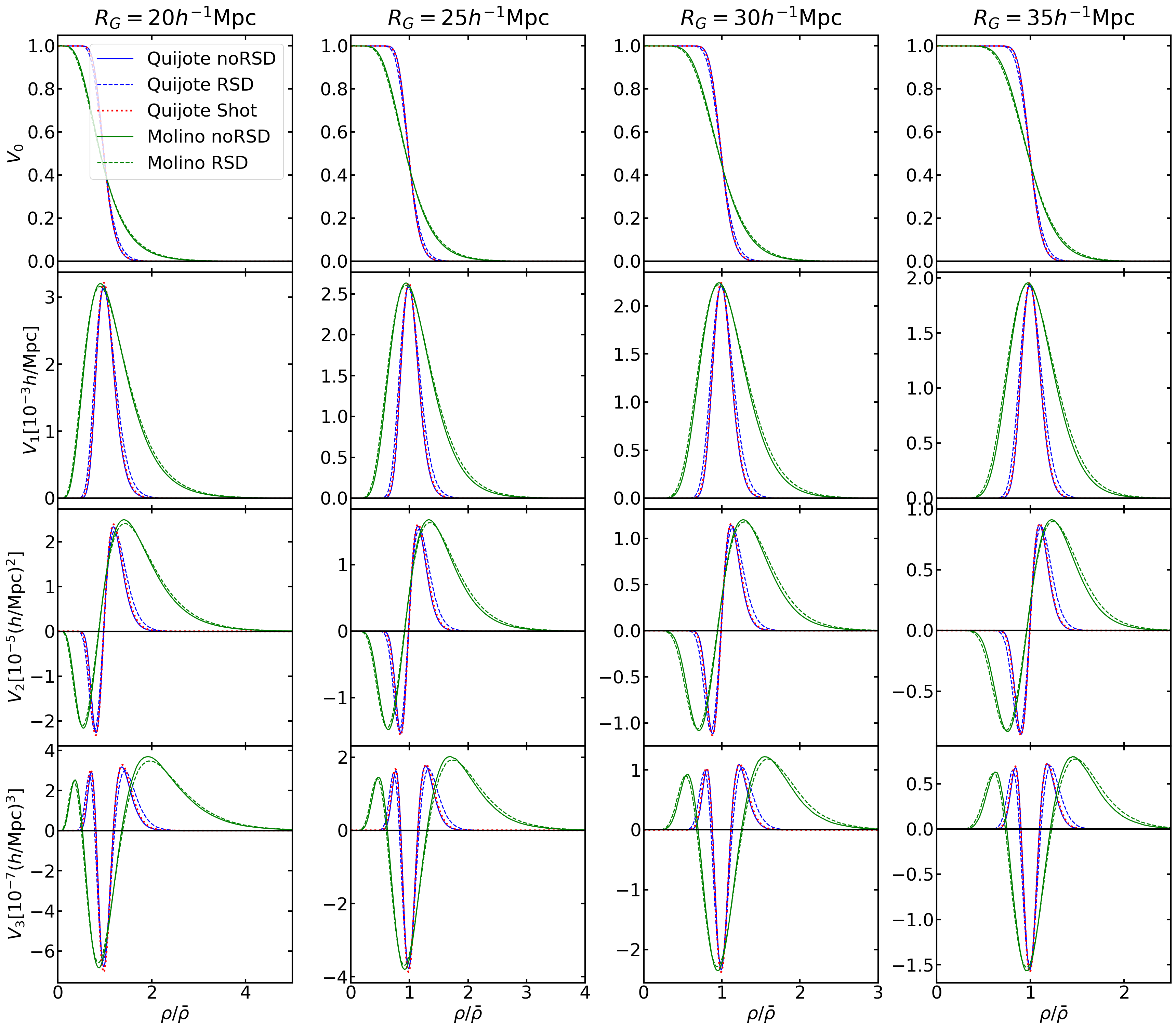}
	\caption{\label{fig:RSD_Shot_Bias} The MFs for the cold dark matter particle distribution of the Quijote simulation at fiducial cosmology in real (labeled as `Quijote noRSD', blue solid lines) and redshift space (labeled as `Quijote RSD', blue dashed lines). To show the effect of shot noise on the MFs, the cold dark matter particles are downsampled so that the number density $\bar{n}\simeq 0.00106$, whose MFs in real space are plotted with red dotted lines and labeled as `Quijote Shot'.  The MFs of the Molino mock galaxy catalogs at fiducial cosmology and HOD model are also displayed to show how galaxy bias affects the curves of the MFs in both real (labeled as `Molino noRSD', green solid lines) and redshift space (labeled as `Molino RSD', green dashed lines). The 1st, 2nd, 3rd, and 4th column displays the MFs with four different smoothing scales $R_G=20,\ 25,\ 30,\ 35h^{-1}\rm{Mpc}$, respectively. }
\end{figure}

As can be seen in figure \ref{fig:RSD_Shot_Bias}, RSD not only reduces the amplitude of $V_{1,2,3}$, but also expands the curves of MFs from $\rho/\bar{\rho}\sim 1$ toward both lower and higher ends of thresholds. This is because the variance of the field becomes larger in redshift space, and the MFs as a function of the density threshold parameter $\rho/\bar{\rho}$ become wider as the variance of the field gets larger.  For a more detailed interpretation of the RSD effect on the MFs, we refer the interested reader to \cite{2021arXiv210803851J}.

\subsection{Shot noise}
\label{sec:shot_noise_correction}
To understand the effect of shot noise on the MFs, it will be helpful to take a look at the MFs of a Gaussian random field, whose MFs are analytically known as (for brevity, the expression is given as a function of $\nu$) \cite{10.1143/PTP.76.952}
\begin{equation}
\label{eq:tomita}
V_k(\nu)=\frac{1}{(2 \pi)^{(k+1) / 2}} \frac{\omega_3}{\omega_{3-k} \omega_k}\left(\frac{\sigma_1}{\sqrt{3} \sigma_0}\right)^k e^{-\nu^2 / 2} H_{k-1}(\nu),
\end{equation}
where $\omega_0=1$, $\omega_1=2$, $\omega_2=\pi$, $\omega_3=4\pi/3$, and the parameters $\sigma_j$ are given by 
\begin{equation}
\sigma_j^2=\int_0^\infty \frac{k^2 dk}{2\pi^2}k^{2j}P_L(k)e^{-(kR)^2/2},
\end{equation}
where $P_L(k)$ is the linear power spectrum\footnote{The Quijote simulations team has made the linear power spectra publicly available, and one can find more information at \href{https://quijote-simulations.readthedocs.io/en/latest/Pk.html}{https://quijote-simulations.readthedocs.io/en/latest/Pk.html}} and the Gaussian window function is assumed. $H_{k-1}(\nu)$ is the $(k-1)$th order Hermite polynomials. In the presence of shot noise, 
\begin{equation}
\label{eq:sigma_0_shot}
\hat{\sigma}_0^2=\int_0^\infty \frac{k^2 dk}{2\pi^2}(P_L(k)+\frac{1}{\bar{n}})e^{-(kR)^2/2}=\sigma_0^2+\frac{1}{(4\pi)^{3/2}R^3\bar{n}},
\end{equation}
 \begin{equation}
\hat{\sigma}_1^2=\int_0^\infty \frac{k^4 dk}{2\pi^2}(P_L(k)+\frac{1}{\bar{n}})e^{-(kR)^2/2}=\sigma_1^2+\frac{3}{16\pi^{3/2}R^5\bar{n}}.
\end{equation}
Therefore, shot noise changes the amplitude of the MFs by increasing both $\sigma_0$ and $\sigma_1$. Shot noise can also affect the MFs by changing the skewness, kurtosis, and higher-order parameters\footnote{The definition of the skewness parameters is given in Equation~\ref{eq:skewness_parameters} and the dependence of the MFs on them can be found in Equation~\ref{eq:first_nonGaussian}. See \cite{2020arXiv201200203M} for the definition of kurtosis parameters and the MFs' dependence on them.} of the density field since non-Gaussian corrections of equation~\ref{eq:tomita} depend on these parameters \cite{2003ApJ...584....1M,2020arXiv201104954M}. However, a detailed study of the shot noise's non-Gaussian effect on the MFs is beyond the scope of this work. Now we only focus on the effect of shot noise in the Gaussian limit and we will also derive a Gaussian correction of shot noise on the MFs.

\begin{figure}[tbp]
	\centering 
	\includegraphics[width=1.0\textwidth]{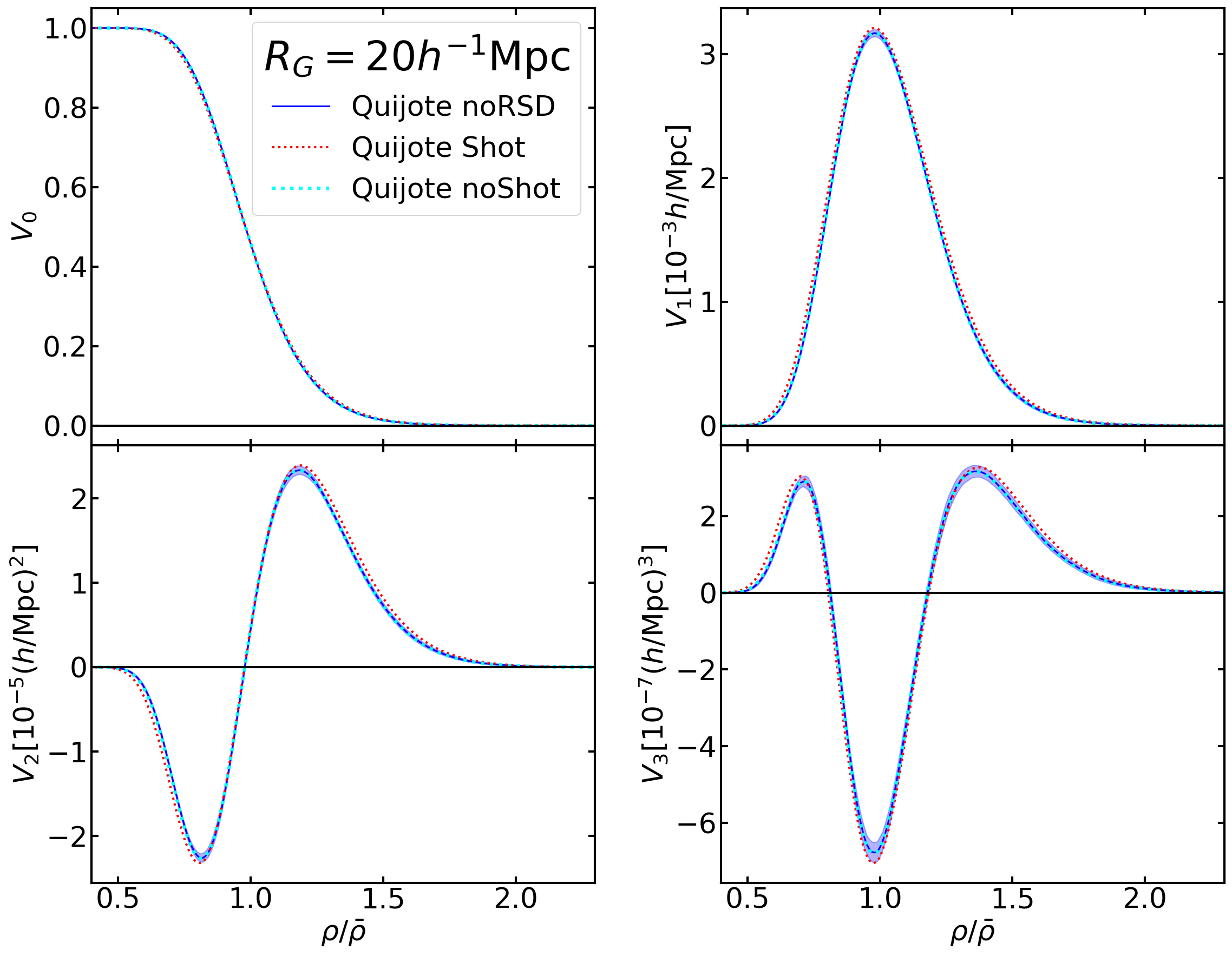}
	\caption{\label{fig:Correct_ShotNoise}
	The MFs for the real-space full (labeled as `Quijote noRSD', blue solid lines) and downsampled (labeled as `Quijote Shot', red dotted lines) CDM particle catalogs of the Quijote simulation. The CDM particles are downsampled to have a number density of $\bar{n}\simeq 0.00106$ for mocking a similar level of shot noise effect existing in the Molino mock galaxy catalogs. We try to correct the effect of shot noise by multiplying the measured MFs with a factor and rescaling the threshold $\rho/\bar{\rho}$. The MFs with shot noise's effect corrected are plotted with cyan dotted lines and labeled with `Quijote noShot'. The corrected MFs lie well within $1\sigma$ error (blue shadow) region of the MFs for full CDM particles (very high particle number density, almost no shot noise for $R_G=20h^{-1}\rm{Mpc}$), which means our correction for the effect of shot noise works well even for $R_G=20h^{-1}\rm{Mpc}$.}
\end{figure}

For the Molino mock galaxies, 
\begin{equation}
(\frac{\hat{\sigma}_0}{\hat{\sigma}_1})_g=\frac{b^2\sigma_0^2+((4\pi)^{3/2}R^3\bar{n_g})^{-1}}{b^2\sigma_1^2+3(16\pi^{3/2}R^5\bar{n_g})^{-1}},
\end{equation}
where $\bar{n_g}$ is the average number density of galaxies, $\bar{n_g}\sim 1.63\times 10^{-4} h^3/\rm{Mpc}^3$, and the bias parameter $b\sim 2.55$. Thus, one can deduce that shot noise's effect on the MFs of the Molino mock galaxies is similar to that of Quijote particle with a number density of $\bar{n}\sim 1.63\times 2.55^2 \times 10^{-4} h^3/\rm{Mpc}^3\sim 0.00106 h^3/\rm{Mpc}^3$. We can compare the MFs of the full CDM particles and those of the downsampled particles, and see how shot noise at this level changes the MFs.  In figure~\ref{fig:Correct_ShotNoise}, we find shot noise increases the amplitude of the MFs and expands the curves toward two ends of $\rho/\bar{\rho}$ because $\hat{\sigma}_0^2=\sigma_0^2+((4\pi)^{3/2}R^3\bar{n})^{-1}>\sigma_0^2$. For $R_G = 20,\ 25 h^{-1}\rm{Mpc}$, shot noise still causes a perceptible impact on the MFs. But with the increase of $R_G$, the effect of shot noise weakens and almost disappears for $R_G\geq 30 h^{-1}\rm{Mpc}$. 

On the other hand, as expected, the effect of shot noise can be corrected by multiplying the measured MFs $V_i$ with $(\frac{\sigma_0\hat{\sigma}_1}{\sigma_1\hat{\sigma}_0})^{i}$, and rescaling the threshold $\rho/\bar{\rho}$ to $\frac{\sigma_0}{\hat{\sigma}_0} \rho/\bar{\rho}-\frac{\sigma_0}{\hat{\sigma}_0} +1$. In figure~\ref{fig:Correct_ShotNoise}, we apply this correction of shot noise to the MFs of the downsampled CDM particles and achieve an excellent agreement with the MFs of the full CDM particles, whose number density is $512^3/1000^3$ and shot noise can be neglected for $R_G=20h^{-1}\rm{Mpc}$. We have tried to correct the effect of shot noise on the MFs of the Molino mock galaxy catalogs, however, we find this correction does not significantly influence our Fisher forecasts. This is understandable: shot noise's effect is very close for different cosmological models and HOD models, thus it is almost canceled out during the process of evaluating the Fisher matrix; the density threshold $\rho/\bar{\rho}$ is rescaled in a similar way for all the MFs present in the estimator of the derivative and covariance matrix, and the amplitude change of the derivative is normalized by the inverse of the covariance matrix (equation~\ref{eq:fisher_mat}). Therefore, we only present the results with the MFs without shot noise correction in this work.

\subsection{Galaxy bias}
%Add some reviews?
Although we mainly focus on the redshift space galaxy distribution in this work and the more proper analytical expression for the MFs with non-Gaussian corrections is given in \cite{2013MNRAS.435..531C}, we will cite Eq.(133) of \cite{2003ApJ...584....1M} (expression for real space) due to its relatively lower complexity to help understand the effect of bias on the MFs. The formulae of MFs as a function of $\nu$ expand to $\mathcal{O}\left(\sigma_0^2\right)$ is given by
\begin{equation}
\label{eq:first_nonGaussian}
\begin{aligned}
V_k(\nu)= & \frac{1}{(2 \pi)^{(k+1) / 2}} \frac{\omega_3}{\omega_{3-k} \omega_k}\left(\frac{\sigma_1}{\sqrt{3} \sigma_0}\right)^k \\
& \times e^{-\nu^2 / 2}\left\{H_{k-1}(\nu)+\left[\frac{1}{6} S^{(0)} H_{k+2}(\nu)+\frac{k}{3} S^{(1)} H_k(\nu)+\frac{k(k-1)}{6} S^{(2)} H_{k-2}(\nu)\right] \sigma_0\right\},
\end{aligned}
\end{equation}
where $S^{(0)}$, $S^{(1)}$, and $S^{(2)}$ are three skewness parameters,
\begin{equation}
\label{eq:skewness_parameters}
\begin{aligned}
& S^{(0)}=\frac{\left\langle f^3\right\rangle}{\sigma_0^4} \\
& S^{(1)}=-\frac{3}{4} \frac{\left\langle f^2\left(\nabla^2 f\right)\right\rangle}{\sigma_0^2 \sigma_1^2}, \\
& S^{(2)}=-\frac{9}{4} \frac{\left\langle(\nabla f \cdot \nabla f)\left(\nabla^2 f\right)\right\rangle}{\sigma_1^4}, \\
\end{aligned}
\end{equation}
and $f$ is the smoothed density field. For the local, deterministic biasing scheme,
\begin{equation}
\delta_g=b\delta+\frac{b_2}{2}(\delta^2-\langle\delta^2\rangle)+...,
\end{equation}
the biased $\sigma_i$ parameter in the lowest order is 
\begin{equation}
\sigma_{0,g}=b\sigma_0,\  \sigma_{1,g}=b\sigma_1,
\end{equation}
and all skewness parameters are transformed to \cite{2003ApJ...584....1M}
\begin{equation}
S^{(a)}_{g}=\frac{S^{(a)}}{b}+\frac{3b_2}{b^2},\ a=0,\ 1,\ 2.
\end{equation}

With the help of the above equations, we can start to interpret the effect of galaxy bias on the MFs in real space. For the fiducial Molino mock galaxy catalog, $b\sim 2.55$, thus $\sigma_{0,g}\sim 2.55\sigma_0$. The curves of the MFs for CDM particles will be expanded a lot to both lower and higher ends of $\rho/\bar{\rho}$ when transforming to the MFs for mock galaxies. On the other hand, we can find in figure~\ref{fig:RSD_Shot_Bias} larger asymmetries exist in the curves of the MFs for galaxies, for example, the peak in $V_3$ is higher and wider for high $\rho/\bar{\rho}$ than that for low $\rho/\bar{\rho}$. This means that non-Gaussian corrections are more important for the MFs of the Molino galaxies. The amplitude of the departure from Gaussianity depends on $\sigma_0$, this character of the MFs can thus be utilized to constrain $\sigma_0$ \cite{2013MNRAS.435..531C,PhysRevD.85.023011}. We note that equation~\ref{eq:first_nonGaussian} only includes the first-order effects of non-Gaussianity, however, the MFs of the Molino mock galaxies may contain second-order and even higher-order non-Gaussianity. The second-order non-Gaussianity is determined by both the skewness parameters and kurtosis parameters of the density field \cite{2020arXiv201104954M,2020arXiv201200203M}, these parameters are all influenced by galaxy bias \cite{2018PhR...733....1D}. Higher-order non-Gaussianities of the MFs may also depend on galaxy bias. 

In figure~\ref{fig:RSD_Shot_Bias}, the galaxy bias seems to impact the MFs in a similar way for real and redshift space. However, the effect of galaxy bias in redshift space is coupled with the RSD effect and thus differs from that in real space. The anisotropies introduced by RSD will deviate the $\sigma_i$, skewness, kurtosis, and higher-order cumulant-related parameters from their isotropic values. Both the deviation of these parameters and themselves depend on bias parameters, but with different dependencies. We refer the interested read to \cite{2013MNRAS.435..531C} for a detailed discussion of the combined effect of RSD and galaxy bias on these parameters and the MFs.

\section{Fisher matrix formalism}
\label{sec:fisher} 
We use the Fisher information matrix \cite{fisher1936use,tegmark1997karhunen} to quantify the constraining power of the MFs and power spectrum on the cosmological parameters, which is defined as
\begin{equation}
	F_{\alpha \beta}=\left\langle-\frac{\partial^{2} \ln \mathcal{L}}{\partial \theta_{\alpha} \partial \theta_{\beta}}\right\rangle,
\end{equation}
where the likelihood $\mathcal{L}$ is assumed to be Gaussian. In Appendix~\ref{sec:non-Gaussian}, we check that the likelihood for the power spectrum and the MFs with $R_G < 40 h^{-1} \rm{Mpc}$ can indeed be approximated by Gaussian. Therefore, it may not be necessary to remove the non-Gaussian components in the data vector, as suggested in \cite{2022arXiv220405435P,2018MNRAS.473.2355S}.  
To avoid introducing artificial information, the covariance matrix of the observables $\boldsymbol{C}$ is assumed to be parameter-independent \cite{2013A&A...551A..88C}. The Fisher matrix can then be written as
\begin{equation}
\label{eq:fisher_mat}
	F_{\alpha \beta}=\frac{\partial \boldsymbol{\mu}}{\partial \theta_{\alpha}}^{\mathrm{T}} \boldsymbol{C}^{-1} \frac{\partial \boldsymbol{\mu}}{\partial \theta_{\beta}},
\end{equation}
where $\boldsymbol{\mu}$ is the theoretical mean for the data vector. It can be the power spectrum, the four MFs, the combination of the power spectrum and MFs, or the combination of MFs with different smoothing scales, etc. $\boldsymbol{C}^{-1}$ is the inverse of the covariance matrix. 

\subsection{Derivatives}
\label{sec:fisher_deri}
For both the cosmological parameters and HOD parameters but except the sum of neutrino masses $M_{\nu}$, the derivatives are estimated as
\begin{equation}
\label{eq:para_deri}
	\frac{\partial \boldsymbol{\mu}}{\partial \theta_{\alpha}}=\frac{\boldsymbol{\mu}(\theta_{\alpha}^{+}) - \boldsymbol{\mu}(\theta_{\alpha}^{-})}{\theta_{\alpha}^{+} - \theta_{\alpha}^{-}},
\end{equation}
where $\boldsymbol{\mu}(\theta_{\alpha}^{+})$ and $\boldsymbol{\mu}(\theta_{\alpha}^{-})$ are estimated as the average of observables from 7500 samples at $\theta_{\alpha}^{+}$ and $\theta_{\alpha}^{-}$, respectively. We check that the 7500 samples are enough to obtain converged parameter constraints in Appendix~\ref{sec:conver}.  For $M_{\nu}$, simulations with $M_{\nu}=0.1,\ 0.2,\ 0.4 \rm eV$ and fiducial cosmology simulation with Zeldovich ICs are available. Thus, three kinds of estimators of the derivative can be constructed
\begin{equation}
\label{eq:nu_deri_1}
	\frac{\partial \boldsymbol{\mu}}{\partial M_{\nu}}=\frac{\boldsymbol{\mu}(dM_{\nu}) - \boldsymbol{\mu}(\theta_{fid}^{ZA})}{dM_{\nu}}+\mathcal{O}\left(d M_{\nu}\right),
\end{equation}

\begin{equation}
\label{eq:nu_deri_2}
	\frac{\partial \boldsymbol{\mu}}{\partial M_{\nu}}=\frac{-2 \boldsymbol{\mu}\left(2 d M_\nu\right)+4 \boldsymbol{\mu}\left(d M_\nu\right)-3 \boldsymbol{\mu}\left(\theta_{fid}^{ZA}\right)}{2 d M_\nu}+\mathcal{O}\left(d M_{\nu}^2\right),
\end{equation}

\begin{equation}
\label{eq:nu_deri_3}
	\frac{\partial \boldsymbol{\mu}}{\partial M_{\nu}}=\frac{\boldsymbol{\mu}\left(4 d M_\nu\right) -12 \boldsymbol{\mu}\left(2 d M_\nu\right)+32 \boldsymbol{\mu}\left(d M_\nu\right)-21 \boldsymbol{\mu}\left(\theta_{fid}^{ZA}\right)}{12 d M_\nu}+\mathcal{O}\left(d M_{\nu}^3\right),
\end{equation}
where $dM_{\nu}$ can be 0.1, 0.2, or 0.4 eV in the first equation, 0.1 or 0.2 eV in the second equation, and 0.1 eV in the last equation. Therefore, a total of 6 different derivatives w.r.t. $M_{\nu}$ with different orders of errors can be obtained using the three estimators. Our results are based on the third one because it is the most accurate numerical derivative estimator. The difference in parameter constraints caused by choices of the estimator and value of $dM_{\nu}$ is discussed in Appendix~\ref{sec:derivative_test}.
For the third estimator, $dM_{\nu}=0.1$eV, $\boldsymbol{\mu}(4dM_{\nu})$, $\boldsymbol{\mu}(2dM_{\nu})$, $\boldsymbol{\mu}(dM_{\nu})$, and $\boldsymbol{\mu}(\theta_{fid}^{ZA})$ are thus estimated as the average of observables from 7500 samples at the $M_{\nu}=0.4,\ 0.2,\ 0.1eV$ cosmologies, and the fiducial cosmology with Zeldovich ICs, respectively. 

\begin{figure}[tbp]
	\centering 
	\subfloat[Derivatives w.r.t. the cosmological parameters]{%
	\includegraphics[width=1.0\textwidth]{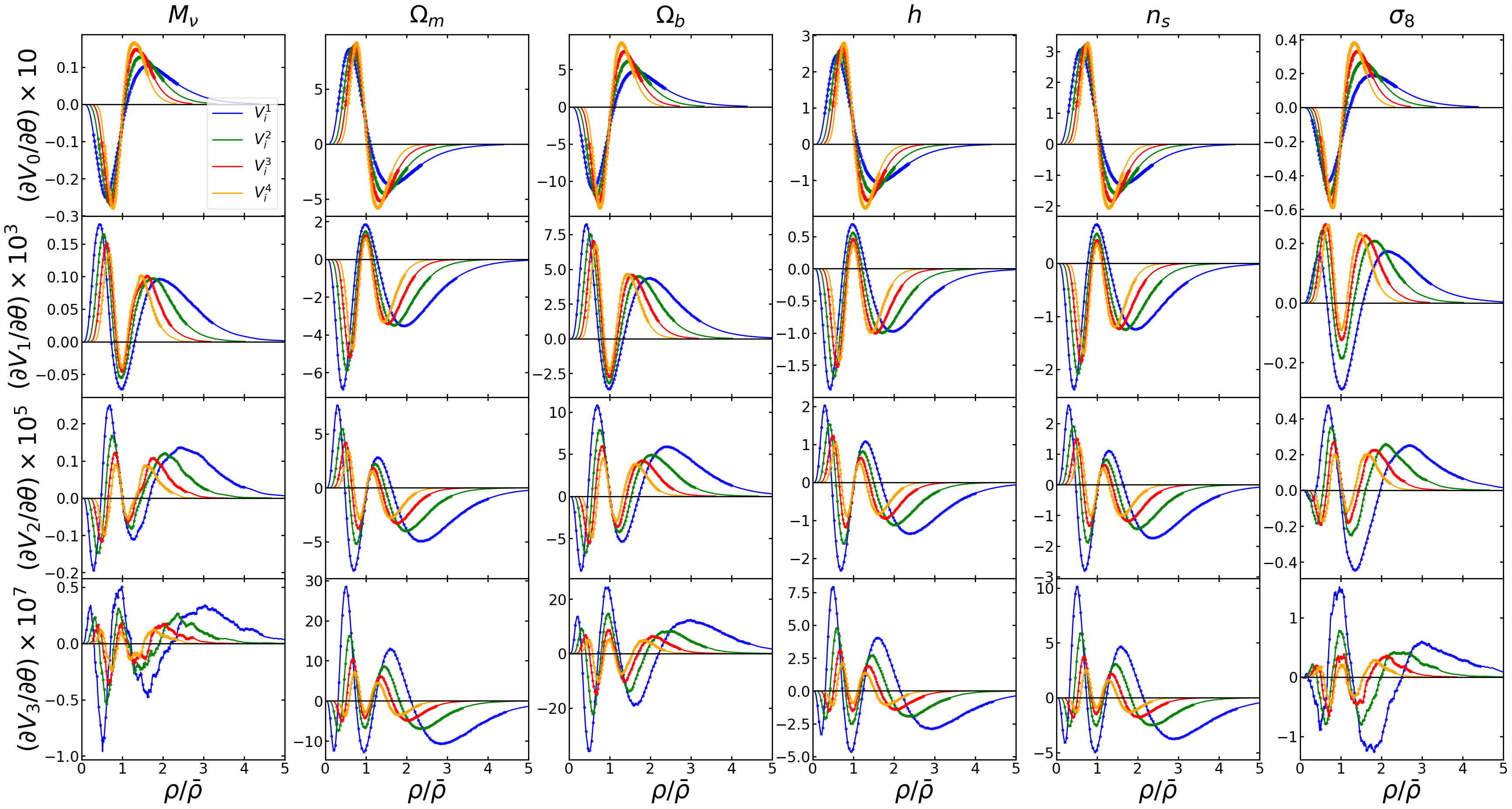}%
	}
	
	\subfloat[Derivatives w.r.t. the HOD parameters]{%
	\includegraphics[width=1.0\textwidth]{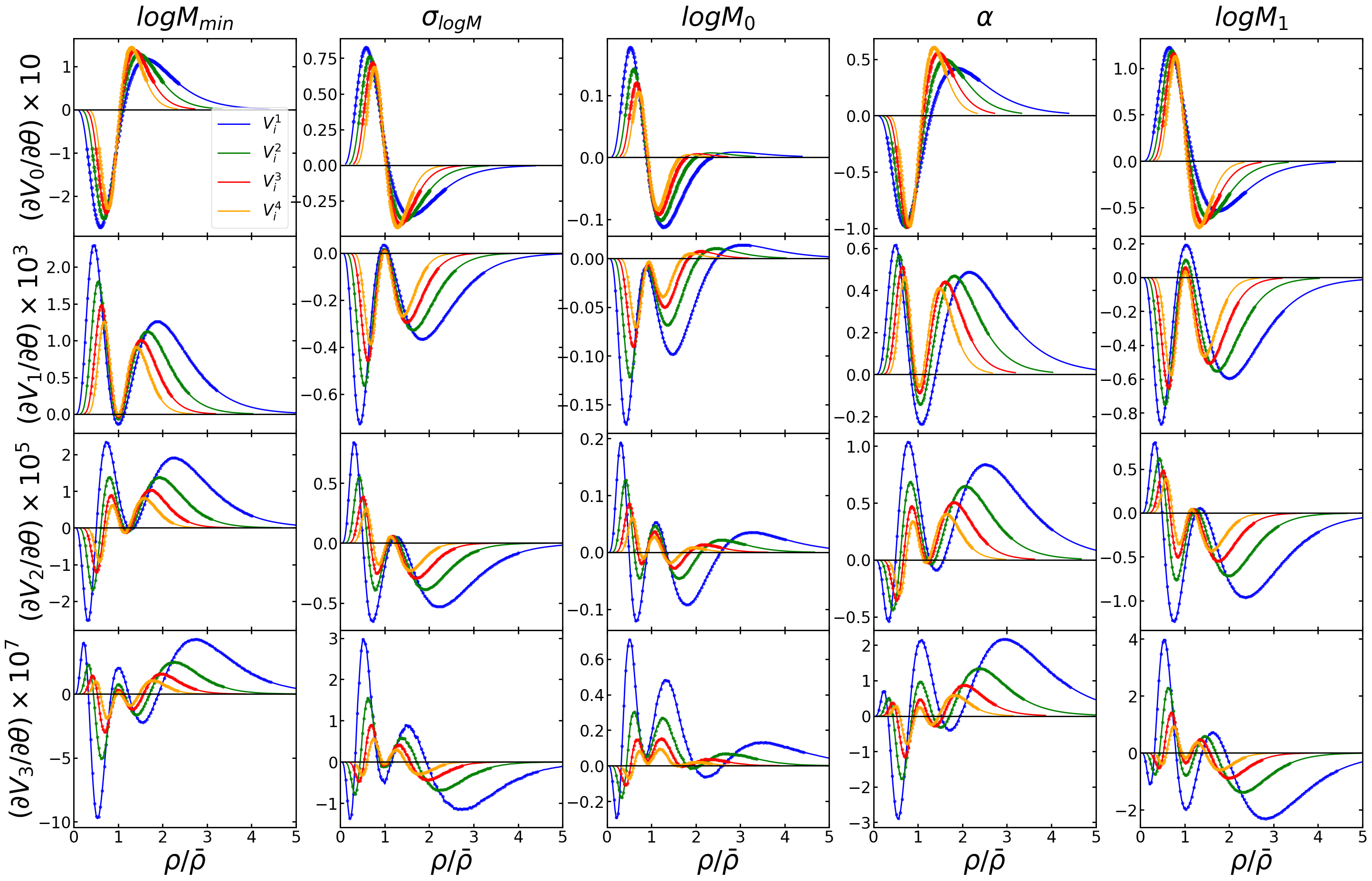}%
	}
	\caption{\label{fig:deri_cos_hod}(a) The numerical derivatives of the four MFs with respect to the six cosmological parameters: $M_{\nu}$, $\Omega_m$, $\Omega_{cb}$, $h$, $n_s$ and $\sigma_8$. (b) The numerical derivatives of the four MFs with respect to the five HOD parameters: $logM_{min}$, $\sigma_{logM}$, $logM_{0}$, $\alpha$, and $logM_1$. For both plots (a) and (b), $V_i^1$, $V_i^2$, $V_i^3$, and $V_i^4$ denote the derivatives of the MFs with $R_G=20,\ 25,\ 30,\ \rm{and}\ 35  h^{-1}\rm{Mpc}$, they are plotted in blue, green, red, and orange, respectively. Points denote the data bins used in the Fisher forecast. See the text for more details about our binning scheme.}
\end{figure}

For the data vector of the MFs, we use a more sophisticated density threshold binning scheme than that used in our previous work \cite{2022JCAP...07..045L}, where we evenly divided the threshold range into $N_{b}$ bins, the two ends of which correspond to the volume fraction of 0.02 and 0.98, respectively. And the four MFs share the same density threshold bins. This old binning scheme ignores the fact that different orders of the MFs have varied threshold ranges where they have non-negligible signals. Our new binning scheme is adaptive for each order of the MFs and aims to solve this issue. \textcolor{black}{We note that the new scheme is based on the mean of MFs at the fiducial cosmology, which is estimated from a total of 15000 realizations. That is, $V_i$ denotes the mean of $V_i$ at the fiducial cosmology in this paragraph.} For $V_0$, we follow our old binning scheme and take $N_b$ evenly spaced threshold bins in the range between $\rho_{0.02}/\bar{\rho}$ to $\rho_{0.98}/\bar{\rho}$, where the thresholds $\rho_{0.02}/\bar{\rho}$ and $\rho_{0.98}/\bar{\rho}$ correspond to the volume fraction of 0.02 and 0.98, respectively. For $V_{1,2,3}$, we can see from figure~\ref{fig:Vis} that $V_{1,2,3}$ are still very close to the peaks and/or valleys of the curves at $\rho_{0.02}/\bar{\rho}$ or $\rho_{0.98}/\bar{\rho}$ and they will approach zero more if the thresholds are increased or reduced further. There is no appropriate universal low and high threshold cut-off for $V_{1,2,3}$. Thus, for each of $V_{1,2,3}$, we first find the lower end of thresholds approximately corresponding to $2\%$ of the maximum of $|V_i|$, $\rho_{low}/\bar{\rho}$; and then find the higher end of thresholds where $|V_i|$ is close to $2\%$ of the maximum of $|V_i|$ as well, $\rho_{\rm{high}}/\bar{\rho}$. Finally, $N_b$ evenly spaced threshold bins are taken in the range between $\rho_{low}/\bar{\rho}$ to $\rho_{high}/\bar{\rho}$. We compare the two threshold binning schemes in figure~\ref{fig:Vis}, which manifests our motivation for this new scheme. The selected data bins are also marked with small points in figure~\ref{fig:deri_cos_hod} for the derivatives of the MFs w.r.t. the cosmological and HOD parameters. Our new threshold binning scheme tries to cover the variation range for each order of the MFs as extensively as possible while avoiding including the bins susceptible to noises and systematics. We find the MFs of too low or high (lower than $\rho_{low}/\bar{\rho}$ or higher than $\rho_{\rm{high}}/\bar{\rho}$) threshold bins may have a non-Gaussian likelihood, hence these bins are not included in our data vector.

For both the monopole and quadrupole of the power spectrum, 79 wavenumber bins are used, up to $k_{\rm max}=0.5h\mathrm{Mpc}^{-1}$. The size of each bin is $2 \pi/L$, where $L=1h^{-1} \rm Gpc$ is the size of the simulation box. We take the same choice of $k_{\rm max}$ as previous works that used the Molino catalogs for Fisher matrix forecasts \cite{2021JCAP...04..029H,2022arXiv220205248W,2022arXiv220601709M,2022arXiv221012743H}, hence the constraints from the power spectrum may serve as a benchmark for the comparison of the MFs with other statistics.

In figure~\ref{fig:deri_cos_hod}, we show the derivatives of the four MFs w.r.t. the cosmological parameters (plot a) and the HOD parameters (plot b) with four different smoothing scales $R_G=$ 20 (blue), 25 (green), 30 (red), and 35 (orange) $h^{-1}\rm{Mpc}$, which are denoted as $\partial V^1_i /\partial \theta$, $\partial V^2_i /\partial \theta$, $\partial V^3_i /\partial \theta$, and $\partial V^4_i /\partial \theta$, respectively. As seen in this figure, the derivatives vary when using different smoothing scales and with respect to different parameters. For the derivative of a particular order of the MFs w.r.t. a specific parameter, the overall shape remains unchanged when smoothing scales vary. However, the curve shifts towards $\rho/\bar{\rho}=1$ from both the low and high threshold sides as the smoothing scale increases, which is caused by the shift of $V_i$ as seen in figure~\ref{fig:Vis}. As for the change of the amplitudes of $\partial V_i /\partial \theta$, it depends on both the variation of the amplitudes of $V_i$ when using different $R_G$s and the scale dependence of the parameters' effects on the LSS. For $\partial V_0 /\partial \theta$, since changing $R_G$ does not alter the amplitude of $V_0$, we do not expect its amplitude to drop with the increase of $R_G$. But for parameters whose effects on the LSS are strongly scale-dependent, like $logM_0$, the amplitude of the $V_0$ derivative falls when larger smoothing scales are used. The overall declining amplitudes of the derivatives for $V_1$, $V_2$, and $V_3$ stem from the amplitude drop of the three Minkowski functionals when increasing the smoothing scale.

For the derivatives with a particular smoothing scale but w.r.t. different parameters, universal patterns exist in each order of the MFs if we change the sign of $\partial V_i / \partial M_{\nu}$, $\partial V_i / \partial \Omega_b$, $\partial V_i / \partial \sigma_8$, $\partial V_i / \partial logM_{min}$, and $\partial V_i / \partial \alpha$. Specifically, one peak and one valley for $\partial V_0 /\partial \theta$, one peak and two valleys for $\partial V_1 /\partial \theta$, two peaks and two valleys for $\partial V_2 /\partial \theta$, and two peaks and three valleys for $\partial V_3 /\partial \theta$. The only exception is $\partial V_{1,2,3} / \partial logM_0$, one more peak is present at the high threshold side in each of the three derivatives. The physical interpretation of these peaks and valleys can be made as in our previous work \cite{2022JCAP...07..045L}, but it will be a tedious job for a total of eleven parameters, thus not given here. Besides the similar overall shapes of $\partial V_i /\partial \theta$, we note that differences exist in the position, amplitude, and steepness of valleys and peaks on the curves of the derivatives for different parameters. This gives us some insights into how the MFs can tightly constrain the cosmological parameters and the HOD parameters at the same time. In addition, how these features evolve when the smoothing scale varies for one parameter is quite distinct from that for another parameter, which indicates that the parameter degeneracies can be broken by combining the MFs with different smoothing scales.

%$\partial V_3 / \partial M_{\nu}$ and $\partial V_3 / \partial \sigma_8$ are a little noisier than the derivatives of $V_3$ w.r.t. other parameters. This is because the effect of changing $M_{\nu}$ or $\sigma_8$ by the step size shown in table~\ref{tab:s} on the MFs of the galaxy field is less significant than changing other parameters. 

%Try to explain where the information of the MFs about all parameters comes from, the number, position, amplitude, and steepness of valleys and peaks.  And how these features change when the smoothing scale varies

%Figure~\ref{fig:deri_cos_hod} also gives us some insights into how the parameter degeneracies are broken when combining the MFs with different smoothing scales. The number, position, amplitude, and steepness of valleys and peaks present on the curves of the derivatives change with respect to both the varied parameters and the differing smoothing scales.

\subsection{Covariance matrix}
\label{sec:fisher_cov}
We estimate the covariance matrices with 15000 HOD catalogs at the fiducial HOD model, which are constructed from the 15000 halo catalogs at the fiducial cosmological model. We check that 15000 independent samples are enough to obtain converged parameter constraints in Appendix~\ref{sec:conver}. Due to uncertainties in the estimated covariance matrix $\hat{\mathbf{C}}$, the inverse of $\hat{\mathbf{C}}$ is not an unbiased estimator for $\mathbf{C}^{-1}$. Following \cite{hartlap2007your}, we remove the bias in $\hat{\mathbf{C}}^{-1}$ by 
\begin{equation}
	\label{eq:debiase}
	\mathbf{C}^{-1}=\frac{n-p-2}{n-1} \hat{\mathbf{C}}^{-1},
\end{equation}
where p is the number of observables in the data vector, and n is the number of simulations used to estimate  $\mathbf{C}$.

\begin{figure}[tbp]
	\centering 
	\includegraphics[width=1.0\textwidth]{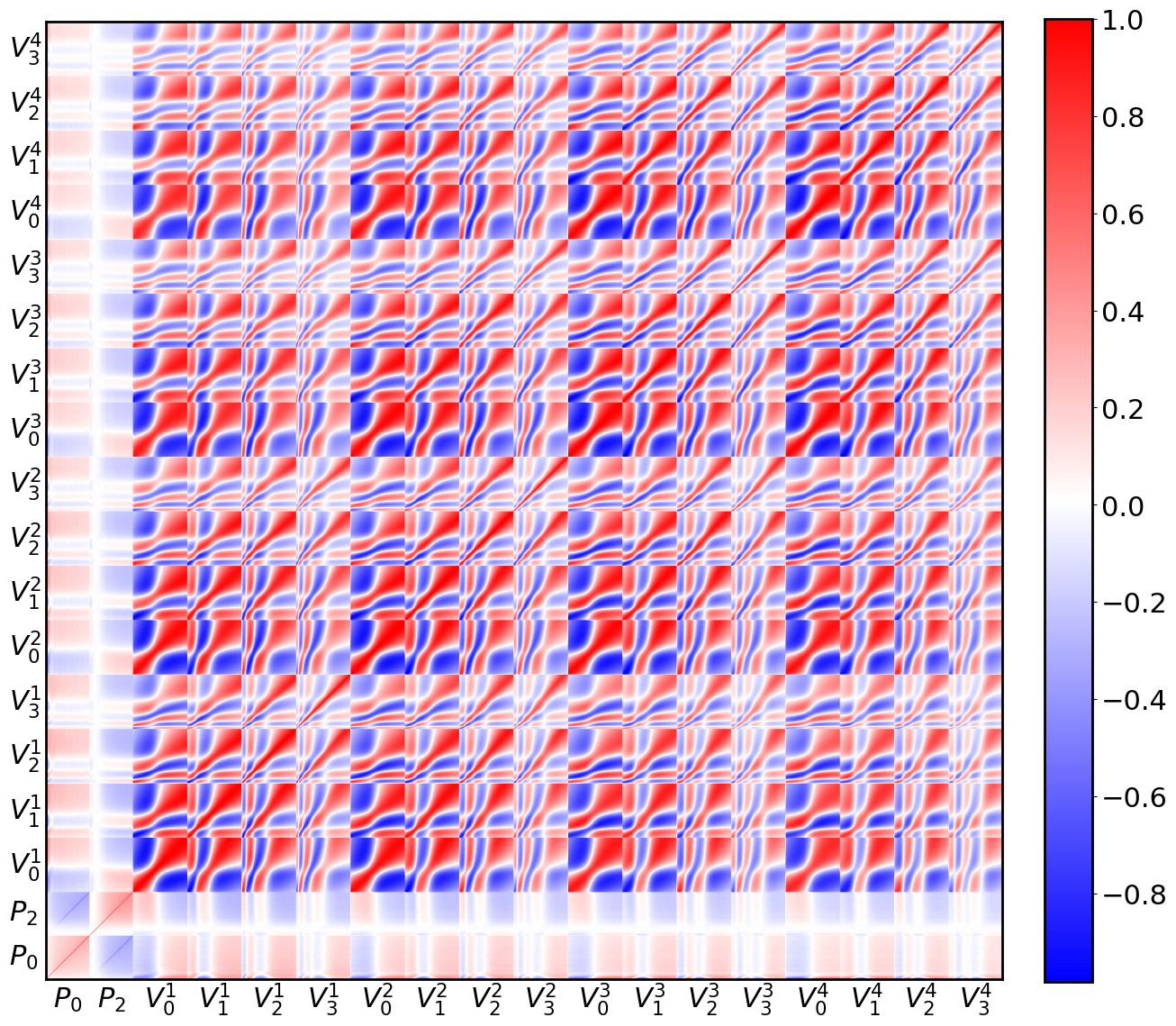}
	\caption{\label{fig:iv} Correlation matrix of the monopole and quadrupole of the power spectrum ($P_{0}\  \rm{and}\ P_2$) and the MFs with four varied smoothing scales (denoted by $V^{1}_i, \ V^{2}_i, \ V^{3}_i, \ \rm{and}\ V^{4}_i$, with the corresponding smoothing scales $R_{G}=20,\ 25,\ 30,\ \rm{and}\ 35\ h^{-1} \mathrm{Mpc}$). For each of the monopole and quadrupole of the power spectrum, 79 wavenumber bins are used, up to $k_{max}=0.5h\mathrm{Mpc}^{-1}$. For each $V^{k}_i$, 100 density threshold bins are selected according to the threshold binning scheme described in the text; thus, a total of 1600 data points are combined for the MFs.  Bin values increase from left to right for each statistic. The correlation matrix is estimated using 15000 HOD galaxy catalogs at the fiducial model. The corresponding covariance matrix is used for the Fisher matrix forecasts. }
\end{figure}

In figure~\ref{fig:iv}, we show the correlation matrices of the data vector that combines the monopole and quadrupole of the power spectrum and the four MFs with four smoothing scales $R_G=20,\ 25,\ 30,\ 35 h^{-1}\rm{Mpc}$. For both the monopole and quadrupole of the power spectrum, 79 wavenumbers are used, up to $k_{max} = 0.5 h\ Mpc^{-1}$. While for the MFs with four different smoothing scales, 400 threshold bins are used for each of the smoothing scales (100 bins for each order of the four MFs). 

The correlation matrix of the power spectrum is shown in the left-bottom corner of figure~\ref{fig:iv}. The auto-correlation of the monopole and quadrupole of the power spectrum has a simple structure: cross-correlation between different wavenumbers is weak on large scales (small $k$-bins, bottom left), but gradually strengthens on smaller scales, where the correlation matrix becomes non-diagonal.
The cross-correlation between the monopole and quadrupole is negative. The small overall correlations between the power spectrum and the MFs suggest that the MFs can provide complementary information, and more stringent constraints can be obtained with the combination of these two statistics \cite{2012PhRvD..85j3513K,2019JCAP...06..019M}. The correlation matrix of the MFs exhibits a much richer structure. There are correlations between the MFs with different threshold bins, orders, and smoothing scales. The MFs at neighboring threshold bins are positively correlated because the excursion sets for neighboring bins have close morphological properties. Different orders of the MFs are also anticipated to be correlated with each other. For example, when density thresholds are high enough so that the excursion sets are mainly isolated high-density islands, the excursion sets with higher thresholds typically have smaller total volume, smaller surface area, smaller integrated mean curvature, and smaller Euler characteristic. As is shown in figure~\ref{fig:Vis}, the MFs with different smoothing scales share similar shapes and features. Hence we expect the MFs with $R_G=20,\ 25,\ 30,\ 35 h^{-1}\rm Mpc$ are also correlated.

When the Fisher matrix is calculated using the simulation-based method, a large number of samples are needed for an accurate estimate of both the derivatives and covariance matrix to obtain convergent parameter constraints \cite{2021arXiv210100298B,2020JCAP...03..040H,2022arXiv220601619C}. In Appendix~\ref{sec:conver}, we show that the number of realizations we have used to estimate the derivatives and covariance matrices is large enough to obtain convergent parameter constraints.

%stop here
\section{Results}
\label{sec:result}
\begin{center}
	\small
	\begin{table}[tbp]
		\scalebox{0.75}[1]{
			\begin{tabular}{|c|c|c|c|c|c|c|c|c|c|}
				\hline Parameters & $P_0+P_2$ & $\Sigma_i V^1_{i}$ & $\Sigma_i V^2_i$ & $\Sigma_i V^3_i$ & $\Sigma_i V^4_i$ & $\Sigma_k \Sigma_i V^k_i$ & $P+\Sigma_k \Sigma_i V^k_i$ & $P/(\Sigma_k \Sigma_i V^k_i)$ & $P/(P+\Sigma_k \Sigma_i V^k_i)$  \\
				\hline
				$M_{\nu}$           & 0.335 & 0.118 & 0.121 & 0.120 & 0.119 & 0.059 & 0.058 & 5.7 & 5.8 \\
				$\Omega_{m}$    & 0.037 & 0.043 & 0.043 & 0.042 & 0.042 & 0.020 & 0.014 & 1.9 & 2.6\\
				$\Omega_{b}$     & 0.0152 & 0.0113 & 0.0111 & 0.0111 & 0.0112 & 0.0053 & 0.0047 & 2.9 & 3.2\\
				$h$                    & 0.178 & 0.092 & 0.092 & 0.093 & 0.108 & 0.048 & 0.042 & 3.7 & 4.2\\
				$n_{s}$               & 0.205 & 0.102 & 0.095 & 0.105 & 0.111 & 0.049 & 0.039 & 4.2 & 5.3\\
				$\sigma_{8}$      & 0.089 & 0.068 & 0.078 & 0.074 & 0.083 & 0.036 & 0.014 & 2.5 & 6.4\\
				\hline
				$logM_{min}$      & 0.457 & 0.089 & 0.110 & 0.141 & 0.175 & 0.045 & 0.023 & 10.2 & 20.0\\
				$\sigma_{logM}$ & 0.963 & 0.136 & 0.143 & 0.142 & 0.158 & 0.071 & 0.064 & 13.6 & 15.0\\
				$logM_0$           & 0.547 & 0.220 & 0.377 & 0.551 & 0.683 & 0.158 & 0.064 & 3.5 & 8.5\\
				$\alpha$            & 0.257 & 0.103 & 0.141 & 0.186 & 0.242 & 0.082 & 0.059 & 3.1 & 4.4\\
				$logM_1$           & 0.193 & 0.135 & 0.192 & 0.291 & 0.413 & 0.088 & 0.070 & 2.2 & 2.8\\
				\hline 
			\end{tabular}
		}
		\caption{\label{tab:error} Marginalized constraints on cosmological and HOD parameters obtained with Fisher matrix analysis for the redshift-space galaxy distribution. They are obtained from the combination of the monopole and quadrupole of the power spectrum ($k_{max}=0.5h\mathrm{Mpc}^{-1}$, $P_0+P_2$, first column); from the four Minkowski functionals with four different smoothing scales: $\Sigma_i V^1_{i}$ with $R_G=20 h^{-1}\rm{Mpc}$ (second column), $\Sigma_i V^2_{i}$ with $R_G=25 h^{-1}\rm{Mpc}$ (third column), $\Sigma_i V^3_{i}$ with $R_G=30 h^{-1}\rm{Mpc}$ (fourth column), and $\Sigma_i V^4_{i}$ with $R_G=35 h^{-1}\rm{Mpc}$ (fifth column); and from the combination of the MFs with the four $R_G$s ($\Sigma_k \Sigma_i V^k_i$, sixth column) as well as the combination of the power spectrum and the MFs ($P+\Sigma_k \Sigma_i V^k_i$, seventh column). The last two columns show the ratios of constraints from the power spectrum to those from $\Sigma_k \Sigma_i V^k_i$ and  $P+\Sigma_k \Sigma_i V^k_i$. Compared to the power spectrum up to quadrupole, the MFs and the combination of the power spectrum and the MFs improve the constraint on $M_{\nu}$ by a factor of 5.7 and 5.8, respectively.} 
	\end{table}
\end{center}

\begin{figure}[tbp]
	\centering 
	\includegraphics[width=1.0\textwidth]{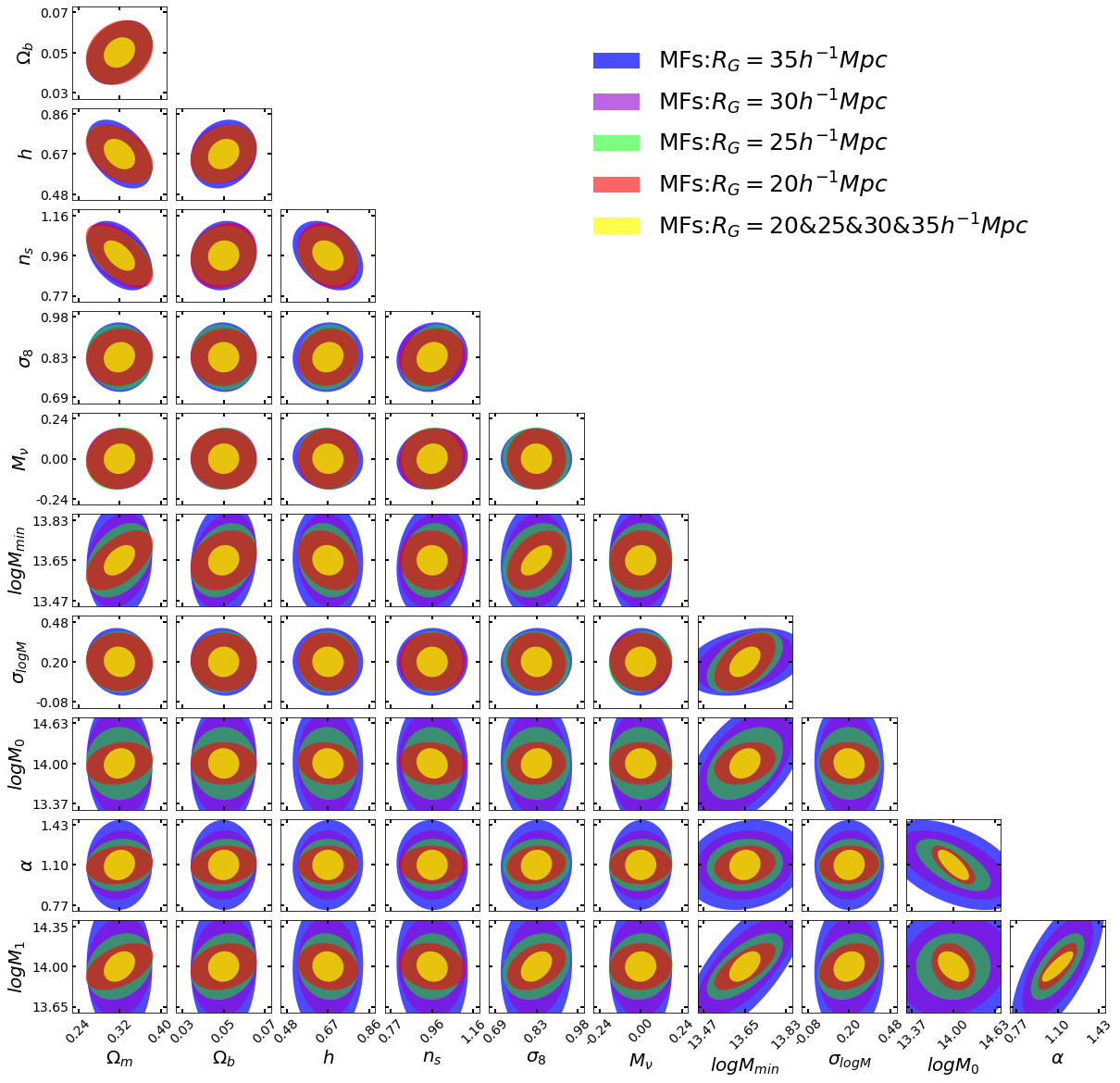}
	\caption{\label{fig:m} The $68\%$ confidence contours for cosmological and HOD parameters from the MFs of the redshift-space galaxy distribution. Blue, purple, green, and red contours for the MFs with the smoothing scale $R_G=35,\ 30,\ 25,\ \rm{and}\ 20\ h^{-1}\mathrm{Mpc}$, respectively. Yellow contours for the combination of the MFs with the four smoothing scales.}
\end{figure}

\begin{figure}[tbp]
	\centering 
	\includegraphics[width=1.0\textwidth]{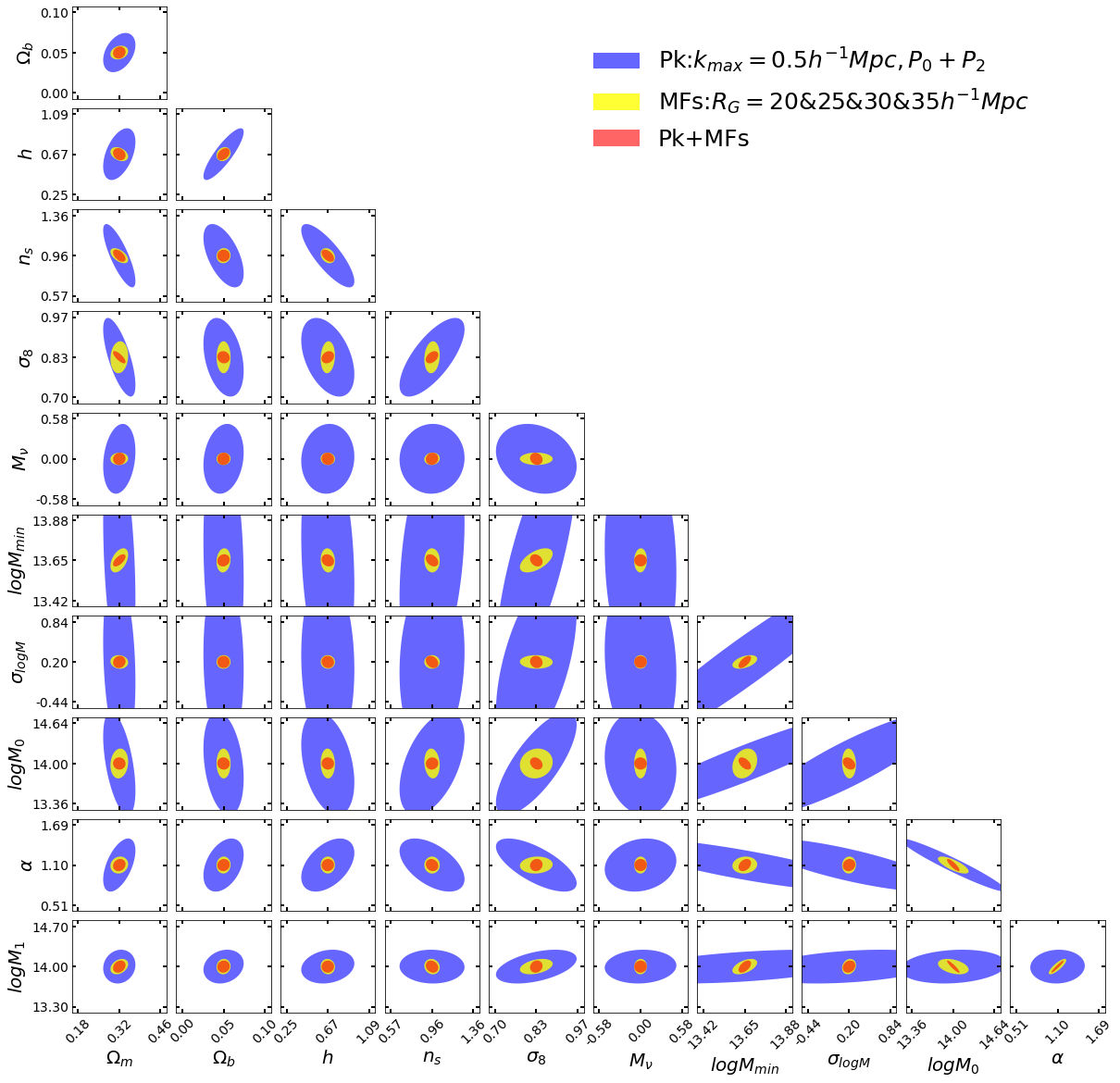}
	\caption{\label{fig:l} The $68\%$ confidence contours for cosmological and HOD parameters from observables for the redshift-space galaxy distribution. Blue contours for the combination of the monopole and quadrupole of the power spectrum with $k_{max}=0.5 h \mathrm{Mpc}^{-1}$, yellow contours for the combination of the MFs with the four smoothing scales $R_G=20\ \&\ 25\ \&\ 30\ \&\ 35 h^{-1}\mathrm{Mpc}$, and red contours for the combination of the power spectrum and the MFs.}
\end{figure}

We show the marginalized $68\%$ Fisher matrix confidence contours obtained from the redshift-space galaxy number density field, first for the MFs with four different smoothing scales and their combination in figure~\ref{fig:m}, and then for the comparison of the power spectrum and the MFs as well as their combination in figure~\ref{fig:l}. The marginalized errors on both cosmological and HOD parameters are listed in table~\ref{tab:error}. Both the marginalized confidence contour and errors are for a $1(\rm{Gpc}/h)^3$ volume.

%The dissimilar parameter degeneracies present in the MFs with differing smoothing scales indicate significant degeneracy breaking and notable constraint improvement when the MFs on the four scales are combined. The combined MFs provide vital complementary information for the power spectrum, parameter degeneracies limiting the constraining power of the power spectrum are broken and the constraints on parameters are significantly improved.                                                         

%The four smoothing scales $R_G=20,\ 25,\ 30,\ \rm{and}\ 35 h^{-1}\mathrm{Mpc}$ are chosen to fulfill the requirements described in section~\ref{sec:mfs}, and $k_{max}$ is set to $0.5 h\ \rm{Mpc}^{-1}$ for both the monopole and the quadrupole of the power spectrum, the same as previous Fisher matrix forecasts based on the Molino catalogs\cite{2021JCAP...04..029H,2022arXiv220205248W,2022arXiv220601709M}. The marginalized errors on both cosmological parameters and HOD parameters are listed in table~\ref{tab:error}. Both the marginalized confidence contour and errors are for a $1(\rm{Gpc}/h)^3$ volume. 

We will start by discussing the dependence of parameter constraints and parameter degeneracies on the smoothing scale $R_G$, because it helps understand the tight constraints obtained from the combination of the MFs with four $R_G$s. And we will describe the confidence contours of the cosmological parameter pairs first, then the pairs with one cosmological and one HOD parameter, and finally, the HOD parameter pairs.

%In fact, more information embedded in the density field is smeared out when larger smoothing scales are applied. If we use as many threshold bins as possible until the information in the MFs is saturated for each smoothing scale, it is anticipated that the constraints on parameters considered in this work deteriorate when $R_G$ is increased. And the number of threshold bins needed to saturate the information content reduces as $R_G$ increases. However, we use an adaptive threshold binning scheme for the MFs with varied smoothing scales, as described in section~\ref{sec:fisher_deri}. This binning scheme aims at extracting as much information as possible for each smoothing scale with the same number of bins. Therefore, it is unsurprising that there is no significant difference between the cosmological constraints from the MFs with the four different smoothing scales, since 100 threshold bins are still not enough to cover all information from the MFs even for $R_G=35 h^{-1}\rm{Mpc}$. On the other hand, 

For the cosmological parameters, weak scale dependence can be seen in parameter constraints and degeneracies from the MFs. 
Although we find the signal-to-noise ratio (SNR) of the MFs is indeed smaller for larger smoothing scales, the correlation between the MFs of different density thresholds, the correlation between different orders of the MFs, and especially, the parameter degeneracies on the MFs together complicate the scale dependence of the constraints from the MFs on the parameters. We have explicitly checked that the MFs with the smaller smoothing scales produce tighter constraints on each parameter when other parameters are fixed, that is, the parameter degeneracy is not considered. Additionally, parameter degeneracies vary notably for differing smoothing scales when considering only a small subset of parameters. However, the full (cosmological plus HOD) parameter degeneracies weaken the scale dependence of both parameter constraints and degeneracies for the cosmological parameters. 

For the contours of the pairs with one cosmological and one HOD parameter, we can see in figure~\ref{fig:m} that they rotate to align with the vertical direction and/or expand in this direction when increasing $R_G$. Exceptions can be seen for the pairs with one cosmological parameter and $\sigma_{logM}$. This is understandable because the effects of $logM_0$, $\alpha$, $logM_1$, and $logM_{min}$ on the LSS are much more sensitive to smoothing scales than $\sigma_{logM}$ and cosmological parameters. $logM_0$, $\alpha$, and $logM_1$ directly control the population of satellite galaxies, they are mainly responsible for structures on small scales, which change greatly when using different smoothing scales. Both $logM_{min}$ and $\sigma_M$ are directly responsible for the population of central galaxies, which mainly influence structures on relatively larger scales. However, the two parameters also indirectly regulate the population of satellite galaxies through equation~\ref{eq:satellite_hod}. Due to the halo mass limit of the Quijote simulation, the fiducial value of $\sigma_{logM}$ is set to 0.2, which is smaller than fits to the SDSS $M_r<-21.5$ and $-22$ sample \cite{Zheng_2007}. When $\sigma_{logM}$ is small, small changes of $logM_{min}$ may cause a big difference in the population of both central and satellite galaxies (see Eq. \ref{eq:central_hod}). Thus $logM_{min}$ may strongly influence small-scale structures as well. %This might explain the contours for $logM_{min}$ and other parameters are also sensitive to the adopted smoothing scale like $logM_0$, $\alpha$, and $logM_1$. 

For the HOD parameters, the confidence contours rotate notably and expand quickly with the increase of $R_G$. This is because HOD parameters mainly influence structure on small scales \cite{2022arXiv220601709M}, as explained above. This strong scale dependence of parameter degeneracies and constraints can also be explained from the perspective of the characteristic scale dependence of the derivatives w.r.t. HOD parameters. The typical scale dependence of the derivatives is distinct from the overall declining amplitude and the shift toward $\rho/\bar{\rho}=1$ from both the low and high ends of the thresholds when the smoothing scale is increased. For example, let's take a close look at $\partial V_3/\partial logM_1$, which is plotted at the right-bottom corner of plot (b) in figure~\ref{fig:deri_cos_hod}. Although the rightmost valley in the curves is always the deepest for all $R_G$s, the ranking of the depth of the valley at low thresholds (the first valley) and the valley at intermediate thresholds (the second valley) is changing with the increase of $R_G$: the first valley is the most shallow one for $R_G=20 h^{-1}\rm{Mpc}$ (blue line); then the two valleys become approximately equally deep for $R_G=25 h^{-1}\rm{Mpc}$ (green line); finally, the second valley become the most shallow one for $R_G=30\ \rm{and}\ 35\  h^{-1}\rm{Mpc}$ (red and orange line). The characteristic scale-dependent features present in the derivatives w.r.t. HOD parameters indicate that distinct information content is embedded in the MFs with different smoothing scales, which leads to scale-dependent parameter constraints and scale-dependent parameter degeneracies. 
%Because of the various scale-dependent degeneracies among the HOD parameters as well as between cosmological parameters and HOD parameters, the constraints on both the cosmological parameters and HOD parameters are strengthened remarkably when the MFs with different $R_G$s are combined.

Although degeneracies among cosmological parameters are similar for the four smoothing scales, they are anticipated to break when the four scales are combined. Because degeneracies between one cosmological parameter and one HOD parameter and those among HOD parameters are quite different for the varied $R_G$s, combining the four scales breaks these degeneracies significantly, which will then help to break degeneracies among cosmological parameters. Therefore, constraints on all the parameters are improved notably. Now that we understand where the constraining power of the MFs comes from, we are prepared to show the MFs provide vital supplementary information for the power spectrum and improve the constraints on parameters significantly.    

We can see in figure~\ref{fig:l} that parameter constraints from the MFs alone are already much tighter than constraints from the power spectrum. With the power spectrum alone, we obtain the $M_{\nu}$, $\Omega_m$, $\Omega_b$, $h$, $n_s$, and $\sigma_8$ constraints: 0.335 eV, 0.037, 0.0152, 0.178, 0.205, and 0.089; while with the MFs alone, we get: 0.059 eV, 0.020, 0.0053, 0.048, 0.049, and 0.036. Compared to the power spectrum, the MFs improve the constraints on $M_{\nu}$, $\Omega_m$, $\Omega_b$, $h$, $n_s$, and $\sigma_8$ by a factor of 5.7, 1.9, 2.9, 3.7, 4.2 and 2.5, respectively. \textcolor{black}{Comparing the correlation coefficients between cosmological parameters from the power spectrum and those from the MFs in Table~\ref{tab:correlation_coefficient}, we find cosmological parameter degeneracies are all weaker for the MFs, with the only exception for the $M_{\nu}-n_s$ degeneracy. The information embedded in the power spectrum only helps reduce degeneracies of $\Omega_m - \Omega_b$, $\Omega_m - \ h$, and $\Omega_b - n_s$. Therefore, although constraints on cosmological parameters from the MFs are all improved when combined with the power spectrum, the improvement of constraining power on $M_{\nu}$ and $\sigma_8$ is not driven by degeneracy breaking.} We note the combination of the MFs and power spectrum only slightly improves the $M_{\nu}$ constraint from the MFs alone, by $2\%$ (from $0.059$ to $0.058$ eV). This finding is consistent with other works based on the Molino mock catalogs: for the bispectrum monopole \cite{2021JCAP...04..029H}, marked power spectrum \cite{2022arXiv220601709M}, and skew spectra \cite{2022arXiv221012743H}, when combined with the power spectrum, the $M_{\nu}$ constraint from the statistics alone is improved by  $3\%$, $9\%$, and $8\%$, respectively. This may indicate the $M_{\nu}$ constraint is dominated by information from higher order statistics.

\begin{center}
	\small
	\begin{table}[tbp]
		\scalebox{0.83}[1]{
			\begin{tabular}{|c|c|c|c|c|c|c|}
				\hline Parameter & $M_{\nu}$ & $\Omega_m$ & $\Omega_b$ & $h$ & $n_s$ & $\sigma_8$ \\
				\hline
				$M_{\nu}$ & 1 & 0.216 & 0.178 & 0.064 & 0.026 & -0.219 \\
				$\Omega_m$ & 0.051 (0.056) & 1 & 0.383 & 0.476 & -0.801 & -0.708 \\
				$\Omega_b$ & 0.018 (0.019) & 0.206 (0.114) & 1 & 0.853 & -0.56 & -0.346 \\
				$h$ & -0.031 (-0.045) & -0.348 (-0.218) & 0.193 (0.38) & 1 & -0.81 & -0.434 \\
				$n_s$ & 0.079 (0.085) & -0.546 (-0.546) & 0.056 (0.025) & -0.26 (-0.487) & 1 & 0.712  \\
				$\sigma_8$ & -0.007 (-0.153) & 0.103 (-0.764) & -0.004 (-0.087) & 0.107 (0.227) & 0.13 (0.396) & 1 \\
				\hline
			\end{tabular}
		}
		\caption{\label{tab:correlation_coefficient} The correlation coefficients between different cosmological parameters for statistics of the redshift-space galaxy distribution. We compare correlation coefficients from the monopole and quadrupole of the power spectrum (upper triangle), the combination of the MFs with four smoothing scales (lower triangle, outside the brackets), and the combination of the power spectra and the MFs (lower triangle, inside the brackets). We can see parameter degeneracies are generally weaker for the Minkowski functionals and the combination of the two statistics.} 
	\end{table}
\end{center}

%For the HOD parameters, the overall improvement of constraints from the MFs is more significant than for the cosmological parameters. Notably, the MFs alone can improve the constraints on the two parameters controlling the number of central galaxies $logM_{min}$ and $\sigma_{logM}$ by a factor of 10.2 and 13.6 over the power spectrum, respectively. This is consistent with our finding of characteristic features present in derivatives w.r.t. the HOD parameters. We also note that the power spectrum's contribution is vital for improving the MF constraints on $logM_{min}$ and $logM_0$. Combined with the power spectrum, the constraints on these two parameters are improved by the MFs from a factor of 10.2 to 20.0 and from 3.5 to 8.5, respectively.

%small scale information is important for constraints on HOD parameters.
%Comparison with constraints in the cold dark matter field
%Comparison with Bispectum
\section{Discussions}
\label{sec:discuss}
The Molino catalogs have been used to quantify information content embedded in the power spectrum and bispectrum \cite{2021JCAP...04..029H}, the combination of power spectra measured from the pre- and post-reconstructed galaxy samples \cite{2022arXiv220205248W}, the marked power spectra of the galaxy field that up-weight low-density regions\cite{2022arXiv220601709M}, and the weighted skew spectra \cite{2022arXiv221012743H}. In section~\ref{sec:compare_statistcs}, we directly compare our results with those obtained with the bispectrum and marked power spectra, and in section~\ref{sec:comp_cdm}, we will present a comparison with the constraints from the cold dark matter obtained in our previous work \cite{2022JCAP...07..045L}.

\subsection{Comparison with results from other statistics using the Molino catalogs}
\label{sec:compare_statistcs}

\begin{center}
	\small
	\begin{table}[tbp]
		\scalebox{0.83}[1]{
			\begin{tabular}{|c|c|c|c|c|c|c|}
				\hline Parameters & $B_0$ & $\Sigma_i M_i$ & $\Sigma_k \Sigma_i V^k_i$ & $P_0+P_2+B_0$ & $P_0+P_2+\Sigma_i M_i$ & $P_0+P_2+\Sigma_k \Sigma_i V^k_i$ \\
				\hline
				$M_{\nu}$           & 0.073 & 0.12 & 0.059 & 0.071 & 0.11 & 0.058 \\
				$\Omega_{m}$    & 0.018 & 0.016 & 0.020 & 0.013 & 0.015 & 0.014 \\
				$\Omega_{b}$     & 0.006 & 0.007 & 0.005 & 0.005 & 0.006 & 0.005 \\
				$h$                    & 0.052 & 0.07 & 0.048 & 0.047 & 0.07 & 0.042 \\
				$n_{s}$               & 0.053 & 0.06 & 0.049 & 0.049 & 0.06 & 0.039 \\
				$\sigma_{8}$      & 0.034 & 0.017 & 0.036 & 0.021 & 0.015 & 0.014 \\
				\hline
			\end{tabular}
		}
		\caption{\label{tab:compare} Marginalized constraints on cosmological parameters obtained with Fisher matrix analysis for statistics of the redshift-space galaxy distribution. $k_{max}=0.5h\mathrm{Mpc}^{-1}$ for statistics in Fourier space. We compare constraints from the bispectrum monopole $B_0$ (first column) \cite{2021JCAP...04..029H}, the combination of four marked power spectra $\Sigma_i M_i$ (second column) \cite{2022arXiv220601709M}, the combination of the MFs with the four smoothing scales $\Sigma_k \Sigma_i V^k_i$ (third column, our results), and the power spectrum in combination with each of the three statistics: $P_0+P_2+B_0$ (fourth column), $P_0+P_2+\Sigma_i M_i$ (fifth column), and $P_0+P_2+\Sigma_k \Sigma_i V^k_i$ (last column). Among all the statistics considered here, the MFs and their combination with the power spectrum obtain the tightest constraint on $M_{\nu}$.} 
	\end{table}
\end{center}

In table~\ref{tab:compare}, we list the marginalized errors on the six cosmological parameters. The first three columns are obtained from individual statistics: the bispectrum monopole $B_0$ \cite{2021JCAP...04..029H} (first column), the combination of four marked power spectra $\Sigma_i M_i$ \cite{2022arXiv220601709M} (second column), and the combination of the four MFs with four smoothing scales $\Sigma_k \Sigma_i V^k_i$ (third column), respectively. The last three columns show the results from the combination of the power spectrum (monopole plus quadrupole) and the three statistics: $P_0+P_2+B_0$ (fourth column), $P_0+P_2+\Sigma_i M_i$ (fifth column), and $P_0+P_2+\Sigma_k \Sigma_i V^k_i$ (last column).

When considered solely, the bispectrum monopole and the MFs show comparable constraining power on $\Omega_m$, $h$, $n_s$, and $\sigma_8$, the relative difference is within $\sim 10\%$. However, the MFs provide more stringent ($\gtrsim 1.2$ times) constraints on $M_{\nu}$ and $\Omega_b$. This is consistent with our expectations. Because the MFs in redshift space can probe modes from the anisotropic bispectrum and the higher-order multispectrum \cite{2013MNRAS.435..531C}: the MFs in redshift space can be expanded in terms of Hermite polynomials and cumulants of the joint probability density function of the density field and its derivatives, where third-order cumulants can be expressed as a sum over the anisotropic bispectrum with different weights, while higher order cumulants will involve $k$ integrals of multispectra. Although adding the bispectrum anisotropic signal is beneficial for parameter constraints \cite{2020JCAP...06..041G}, the anisotropic signal can be mixed with the isotropic one for the MFs \cite{2013MNRAS.435..531C}. In addition, we find there is no significant difference between the constraints from the real- and redshift-space MFs. Hence we conclude the relatively tighter constraints on $M_{\nu}$ and $\Omega_b$ may come from the high-order multispectrum (beyond bispectrum) information embedded in the MFs.   
 
%\cite{2020JCAP...06..041G}, Minkowski functional in redshift space contain information beyond bispectrum monopole
%The length of our observables is comparable with Hahn 2021
%The MFs contain information from high-order statistics
%The MFs are easier to measure?
%Model of the covariance matrix of the MFs?

Compared with the marked power spectra, the MFs provide tighter constraints on $M_{\nu}$, $\Omega_b$, $h$, and $n_s$, while they are less sensitive to $\Omega_m$ and $\sigma_8$. The constraining power of the marked power spectra up-weighting low-density regions on $M_{\nu}$ is limited because the galaxies in the Molino mock catalogs mainly live in high-density regions; therefore, the low-density regions, where the massive neutrinos' effect is most prominent, might not be accessible for the galaxy marked power spectra \cite{2022arXiv220601709M}. The constraint on $M_{\nu}$ from the MFs is also limited by the same factor. This effect is important for the MFs with small density thresholds. However, massive neutrinos' effects can be seen in the MFs of intermediate and high-density thresholds, and a large amount of information still exists in the MFs of these thresholds \cite{2022JCAP...07..045L}. For constraint on $\sigma_8$, it is found in \cite{2022arXiv220601709M} that essential information exists in the quadrupole of the power spectrum and the marked power spectrum. Due to the rotational invariance of the MFs, they are not sensitive to direction-dependent signals in the density field caused by the redshift-space distortion effect, which may explain the relatively weak constraint on $\sigma_8$ from the MFs. Hence we anticipate improvement in the constraints from the MFs when the lost direction-dependent information is added. For example, we find $\sigma_8$ is better constrained when combining the MFs with $P_2$ than $P_0$ because of the anisotropic information embedded in the quadrupole of the power spectrum. An alternative method to recover the direction-dependent information is to apply the Minkowski tensors to redshift-space galaxy distribution. Unlike the MFs, the Minkowski tensors are a set of functions sensitive to directionally dependent signals in the data \cite{2019ApJ...887..128A,2022arXiv220810164A}, they might be a powerful probe of cosmological parameters and provide extra information for the MFs.

When comparing results from the combined statistics, we find the combination of the power spectrum and the MFs provide the tightest constraints on $M_{\nu}$, $\Omega_b$, $h$, $n_s$ and $\sigma_8$, while the tightest constraints on $\Omega_m$ come from the combination of the power spectrum and the bispectrum. For $M_{\nu}$, the improvement from adding the power spectrum is slight for all three statistics. For $n_s$ and $\sigma_8$, although the MFs and the bispectrum individually put close constraints on the two parameters, the addition of the information in the power spectrum is more helpful in improving constraints on the two parameters for the MFs than the bispectrum.

\subsection{Comparison with results for the cold dark matter field}
\label{sec:comp_cdm}

%the reason for the difference: 1, HOD parameters weaken cosmological constraints; 2, Molino catalog galaxies do not trace under dense regions well;
% explain why the improvement of the constraint on Omega_m from the MFs is 0.9

\begin{center}
	\small
	\begin{table}[tbp]
		\scalebox{1.0}[1]{
			\begin{tabular}{|c|c|c|c|c|c|c|}
        			  \hline \text { Parameters } & $P_{\mathrm{cb}}$ & $\mathrm{MF}_{\mathrm{cb}}$ & $P_{\rm{cb}}/\mathrm{MF}_{\mathrm{cb}}$ & $P_{0+2}$ & $\Sigma_i V_i^1$ & $P_{0+2}/\Sigma_i V_i^1$ \\
                                \hline $M_\nu$ & 1.194 & 0.061 & 19.6 & 0.335 & 0.118 & 2.8 \\
                                $\Omega_m$ & 0.040 & 0.026 & 1.5 & 0.037 & 0.043 & 0.9 \\
                                $\Omega_b$ & 0.015 & 0.006 & 2.5 & 0.0152 & 0.0113 & 1.3 \\
                                $h$ & 0.147 & 0.086 & 1.7 & 0.178 & 0.092 & 1.9 \\
                                $n_s$ & 0.094 & 0.093 & 1.0 & 0.205 & 0.102 & 2.0 \\
                                $\sigma_8$ & 0.068 & 0.013 & 5.2 & 0.089 & 0.068 & 1.3 \\
                                \hline
			\end{tabular}
		}
		\caption{\label{tab:compare_MFnu} Marginalized errors and their ratios of cosmological parameters obtained with Fisher matrix analysis for statistics of the real-space cold dark matter plus baryon distribution (the first three columns) and the redshift-space galaxy distribution (the last three columns). For both the real-space `cb' field and the redshift-space galaxy field, we show constraints from the power spectrum with $k_{max}=0.5h/\rm{Mpc}$ (first column for $P_{cb}$ and fourth column for $P_{0+2}$), the Minkowski functionals with $R_G=20h^{-1}\rm{Mpc}$ (second column for $\rm{MF}_{cb}$ and fifth column for $\Sigma_i V_i^1$), and the ratio of the constraints from the power spectrum to those from the MFs (third column for $P_{\rm{cb}}/\mathrm{MF}_{\mathrm{cb}}$ and last column for $P_{0+2}/\Sigma_i V_i^1$). } 
	\end{table}
\end{center}

In our previous work \cite{2022JCAP...07..045L}, we performed a similar Fisher matrix analysis for the MFs of the real-space cold dark matter plus baryons distribution (the `cb' field). However, the constraints are obtained with a conservative threshold bin choice: 20 threshold bins for each of the four MFs and with a conservative estimate of the derivative w.r.t. $M_{\nu}$ (the estimator Eq.(\ref{eq:nu_deri_1}) is used in our previous work). In this work, we calculate the constraints from the MFs of the `cb' field with a more extensive bin choice: 100 threshold bins for each of the MFs and use Eq.(\ref{eq:nu_deri_3}) to estimate the derivative w.r.t. $M_{\nu}$, for a more straightforward comparison with the constraints from the MFs of the redshift-space galaxy field. 

We present constraints on the cosmological parameters from the power spectrum and the MFs for both fields and the ratio of constraints from the two statistics in table~\ref{tab:compare_MFnu}.  Different from the power spectrum, for which the constraints on $M_{\nu}$ and $\Omega_m$ are tighter in the galaxy field, the constraints on all cosmological parameters from the MFs deteriorate when going from the real-space `cb' field to the redshift-space galaxy field. Focusing on the constraint ratios from the two statistics, we note the ratios are all smaller for the redshift-space galaxy field than for the real-space `cb' field, except for $h$ and $n_s$. In particular, the ratios are significantly reduced for $M_{\nu}$ and $\sigma_8$: from 19.6 and 5.2 to 2.8 and 1.3, respectively. This is understandable: the effect of massive neutrinos is more pronounced in low-density regions than in high-density regions \cite{Massara_2015}; however, the galaxies in the Molino catalogs mainly live in high-mass halos, and there are few galaxies in very low-density regions \cite{2022arXiv220601709M}; therefore, the MFs of galaxies cannot access the information embedded in low-density regions while the MFs of the `cb' particles can; without information from low-density regions, the $M_{\nu}-\sigma_8$ degeneracy cannot be fully broken, hence the improvement from the MFs on the constraints of the two parameters is slighter for the galaxy field than for the `cb' field. On the other hand, the MFs' insensitivity to direction-dependent information induced by the RSD effect may further reduce their constraining power on cosmological parameters. At the same time, the quadrupole of the power spectrum is included to extract direction-dependent information in the Fisher analysis for the redshift-space galaxy field. This may also explain the slighter improvement from the MFs for $\Omega_m$ and $\Omega_b$. The statistics more sensitive to anisotropy, such as the Minkowski tensors \cite{2019ApJ...887..128A} and the 2D Minkowski functionals of slices of 3D volume at different angles to the line-of-sight \cite{2013MNRAS.435..531C}, may be needed to extract more information from the redshift-space galaxy distribution. 

\section{Conclusions}
\label{conclusions}
Cosmology provides an independent probe of the mass sum of neutrinos. Although model-dependent, observation of CMB and LSS provides the current tightest constraint on $M_{\nu}$. To give tighter bounds on $M_{\nu}$, we explore the possibility of extracting more information embedded in LSS using summary statistics beyond traditional two-point statistics.

In \cite{2022JCAP...07..045L}, we studied the potential of constraining neutrino mass using the morphological properties of LSS, represented by the four MFs. We found the effects of massive neutrinos on LSS can be well captured by the MFs. The MFs can not only extract information consistent with statistics such as the halo mass function and void size function, but also capture the distinct effects of massive neutrinos on intermediate density regions with $\rho/\bar{\rho}$ around $1$, in addition to providing extra topological information. Hence the MFs provide amounts of information and place tight constraints on $M_{\nu}$, $\sigma_8$, and other cosmological parameters. However, the results in \cite{2022JCAP...07..045L} are obtained from the distribution of the simulated CDM particles for the `cb' density field and also massive neutrino particles for the `m' density field. In reality, the `m' density field cannot be observed directly, although it gives much tighter constraints than the `cb' field. Generally speaking, weak lensing can probe the projected 2D `m' field, but efforts are still needed to reconstruct the 3D `m' field from it to allow the 3D MFs measurable. As for the `cb' field, galaxy and other surveys can trace it, but are subject to tracer biases and systematic effects such as redshift-space distortions, the Alcock-Paczynski effects, etc.

In this work, we extend the forecasts for the MFs using the Molino mock galaxy catalogs to include a realistic galaxy bias model in redshift space. For the first time, we quantify the information content and constraining power of the MFs on $M_{\nu}$ for the redshift-space galaxy field. Using the combination of the MFs with four varied smoothing scales, the constraints on $M_{\nu}$, $\Omega_m$, $\Omega_b$, $h$, $n_s$, and $\sigma_8$ are improved by a factor of 5.7, 1.9, 2.9, 3.7, 4.2 and 2.5, respectively, compared with the power spectrum. When the power spectrum is added, parameter degeneracies are broken further, and the power spectrum constraints on $M_{\nu}$, $\Omega_m$, $\Omega_b$, $h$, $n_s$, and $\sigma_8$ are further improved, by a factor of 5.8, 2.6, 3.2, 4.2, 5.3 and 6.4, respectively. With a volume of $1(h^{-1}\rm{Gpc})^3$, we can achieve a 0.059 eV or 0.058 eV constraint on $M_{\nu}$ using the MFs alone or combined with the power spectrum, after marginalizing over the HOD parameters. These results demonstrate that the MFs are promising and can be applied to upcoming surveys for tighter constraints on $M_{\nu}$ and other cosmological parameters. 

However, our results might be optimistic, and we need to note several potential caveats, such as the shot noise in the redshift-space galaxy field, the non-Gaussianities in the likelihood of the MFs, the noises in the estimated covariance matrix and numerical derivatives, and the uncertainties in the various types of the $M_{\nu}$ derivative estimators. They are discussed detailedly in Section~\ref{sec:shot_noise_correction}, Appendix~\ref{sec:non-Gaussian}, \ref{sec:conver}, and \ref{sec:derivative_test}, respectively. \textcolor{black}{We demonstrate shot noise does not significantly impact our results for the smoothing scales used in this work}, proper smoothing scale range and density threshold range are selected to exclude the non-Gaussianities in the likelihood of the MFs, sufficient catalogs are utilized to obtain converged estimates of the covariance matrix and derivatives, the most accurate estimator is chosen to calculate derivatives w.r.t. $M_{\nu}$. We conclude our constraints are robust, and these caveats only impact our results slightly.

Our constraints on neutrino mass from the MFs can be further improved. First, optimizing binning schemes for the MFs can improve the constraints. Although we have improved our binning scheme of density threshold in this work, there is still considerable room for optimization in this scheme. Next, although combining the MFs with different smoothing scales may not be the most efficient way to extract structure information from multi-scales, it helps to break parameter degeneracies, for the degeneracy directions from MFs with different smoothing scales are usually different. Thus, combining more smoothing scales is expected to enhance the constraints. Then, combining MFs at multiple redshifts can tighten the constraints, too, as suggested by previous studies with weak lensing MFs \cite{2012PhRvD..85j3513K,2019JCAP...06..019M}. Finally, a larger survey volume and denser galaxy population can further reduce errors. Current and upcoming galaxy surveys, such as DESI, Euclid, and CSST, will cover tens of $(h^{-1}\rm Gpc)^3$ and observe galaxies with a larger number density. In~\cite{2023arXiv230504520J}, we find cosmological constraints from the MFs scale with the survey volume $V$ roughly as $1/\sqrt{V}$.  If we forecast for a survey volume like this, the constraints we obtain would be better by a factor $\sim10$. Additionally, the denser galaxy population will enable the MFs to better extract information encoded in low-density regions than the mock galaxy catalogs used in this forecast. Because massive neutrinos' effects are most prominent on voids, which are better traced by denser galaxy catalogs. And the MFs with low density thresholds are very sensitive to the morphological changes of the voids caused by massive neutrinos. On the other hand, galaxy catalogs with a larger number density will also enable the use of smaller smoothing scales. Hence more information will be captured with the MFs. We expect these surveys to have a good chance of detecting masses of neutrinos with a high significance level through measurements of the MFs.

Several systematics should be considered when applying the MFs to real surveys. One has to take care of the irregular shapes of the masked regions and survey boundaries. However, the effects of these systematics plus varying radial and angular selection functions can be corrected, and the MFs for real galaxy catalogs can be unbiasedly reconstructed \cite{2014MNRAS.437.2488B,2022ApJ...928..108A}.   Another possible source of systematic is the baryonic effects, which are important on small scales. It is unclear how baryonic effects change the MFs of LSS and the derived constraints on neutrino mass. However, \cite{Villaescusa_Navarro_2018,2017MNRAS.471..227M} found the impact of baryonic physics and free-streaming of massive neutrinos on LSS can be treated independently. 
In future work, we plan to investigate all these systematic effects comprehensively, and extract the constraints on neutrino mass by measuring the MFs from real surveys.

To apply the MFs to upcoming redshift surveys, one approach is to smooth the galaxy field with a large enough scale and compare the measurements to the analytical predictions \cite{2022ApJ...928..108A}, which are known for Gaussian density fields \cite{Tomita_1990} and weakly non-Gaussian fields \cite{2003ApJ...584....1M,2020arXiv201104954M}. Even though amounts of information will be smeared out due to the large smoothing scales, systematical effects on the MFs are more minor. They can be relatively easily corrected \cite{2014ApJS..212...22K,2022ApJ...928..108A}.  An alternative approach is to smooth the field with small scales and fit the measurements with the simulation-based interpolative surrogate models of the MFs, known as emulators \cite{2010ApJ...715..104H,2017ApJ...847...50L,2022arXiv220712345M}. To construct an emulator for the MFs, $\sim 100$ sampling points in the cosmological parameter space are needed to model their cosmological dependence, and $\sim 50$ HOD models for each cosmology model are required to model galaxy bias accurately \cite{2019ApJ...874...95Z,2022MNRAS.515..871Y,2022arXiv220308999Z}. \textcolor{black}{Alternatively, we can also follow the methodology of \cite{2023arXiv230709134P} to obtain accurate galaxy fields in redshift space by combining a Lagrangian-space bias expansion with displacement and velocity fields measured from N-body simulations.} Proper handling of systematics such as redshift-space distortion, complex survey geometry, and selection functions is also essential to avoid introducing biases into parameter inference using the emulator models.

Recently, emulators of the MFs have been constructed only for the analysis of 2D weak lensing convergence maps \cite{2015PhRvD..91j3511P,2019JCAP...06..019M}. In future work, we plan to build an emulator for the MFs of 3D redshift-space galaxy distribution and extract the constraints on neutrino mass by comparing the measurements from redshift surveys and predictions from the emulator models. \textcolor{black}{For the modeling of the dark matter-galaxy connection, we will still rely on the halo occupation distribution framework. In order to validate the robustness of the HOD approach to model galaxy distribution for the MFs and the inferred cosmological parameter constraints, we will test our emulator against mock galaxy catalogs generated with the SubHalo Abundance Matching method (SHAM) \cite{2018ARA&A..56..435W}, which assign galaxies to dark matter halos by matching the mass or luminosity of a galaxy with the mass or velocity of the dark matter halo. Testing the HOD-based emulator with the SHAM model can tell us whether the parameter constraints are biased by the mock galaxy construction algorithm since the two methods are based on very different assumptions. Similar tests have been done for the emulator of the 2-point correlation function in \cite{2022arXiv220308999Z} and the wavelet scattering transform in \cite{valogannis_wst}. They found that their HOD-based emulators were able to recover the input cosmology successfully (within $1\sigma$ level), and thus validated their emulator construction.}

\acknowledgments
We thank ChangHoon Hahn for the helpful discussions. We are very grateful to the anonymous referee for the constructive comments, which helped to improve the quality of this paper. This work is supported by the National Natural Science Foundation of China Grants No. 12173036 and 11773024, by the National Key R$\&$D Program of China Grant No. 2021YFC2203100 and No. 2022YFF0503404, by the China Manned Space Project “Probing dark energy, modified gravity and cosmic structure formation by CSST multi-cosmological measurements” and Grant No. CMS-CSST-2021-B01, by the Fundamental Research Funds for Central Universities Grants No. WK3440000004 and WK3440000005, by Cyrus Chun Ying Tang Foundations, and by the 111 Project for "Observational and Theoretical Research on Dark Matter and Dark Energy" (B23042).

\appendix

\section{Non-Gaussianity test}
\label{sec:non-Gaussian}
It was reported in \cite{2022arXiv220405435P} that non-Gaussianities in the likelihood of statistics might lead to artificially tight bounds on the cosmological parameters using the Fisher matrix formalism. To access the Gaussianity of the likelihood of the power spectrum,  the combination of the MFs with four smoothing scales, and the combination of the power spectrum and the MFs, we follow the analysis performed in \cite{10.1093/mnras/stab2384,2022arXiv220904310P} and check that the likelihood of these statistics can be approximated by the multivariate Gaussian. Since we have 15000 mock catalogs for the fiducial model, we can obtain 15000 $\chi^2$ values for each of the statistics by 
\begin{equation}
\chi^2_i=(\boldsymbol{d_i}-\boldsymbol{\mu})^T C^{-1} (\boldsymbol{d_i}-\boldsymbol{\mu}),
\end{equation}
where $\boldsymbol{d_i}$ is the data vector of the summary statistics for the $i$-th catalog, $\boldsymbol{\mu}$ and $C$ is the mean and the covariance matrix of the data vector estimated from the 15000 catalogs. 

\begin{figure}[tbp]
	\centering
	\includegraphics[width=1.0\textwidth]{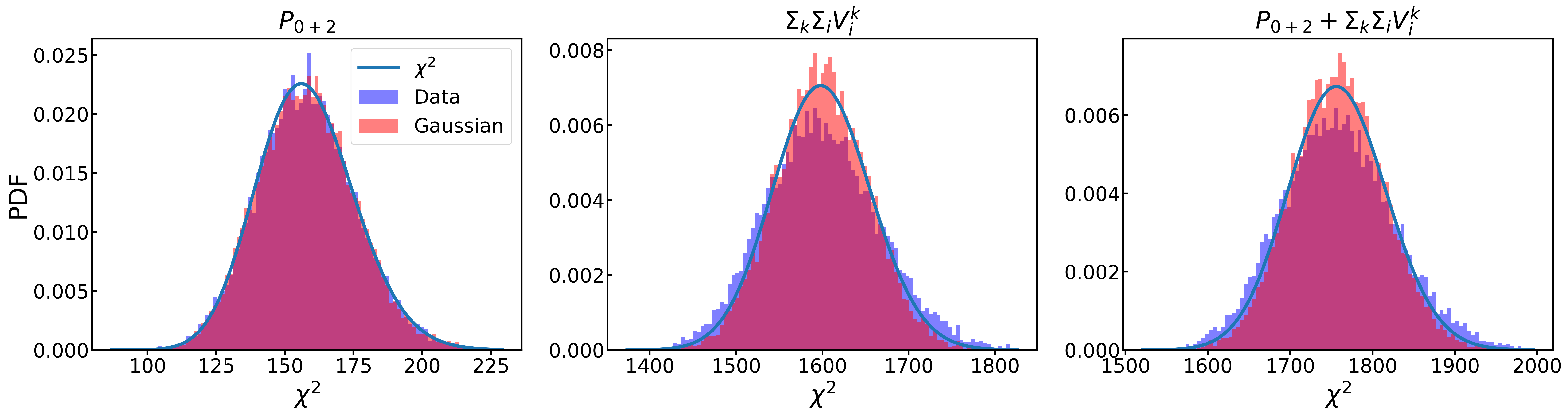}
	\caption{\label{fig:gaussian_test} A qualitative assessment of the Gaussianity of the likelihood for the monopole and quadrupole of the power spectrum ($P_{0+2}$, left column), the combination of the MFs with four smoothing scales ($\Sigma_k\Sigma_i V^k_i$, middle column), and the combination of the power spectrum and the MFs ($P_{0+2}+\Sigma_k\Sigma_i V^k_i$, right column). The histograms of the $\chi^2$ values measured from the Molino catalogs are drawn in blue, while those of the $\chi^2$ values measured from a multivariate Gaussian distribution with the same mean and covariance as the Molino catalogs are shown in red. The solid lines show theoretical $\chi^2$ distributions with degrees of freedom equal to the total number of observables (158 for $P_{0+2}$, 1600 for $\Sigma_k\Sigma_i V^k_i$, and 1758 for $P_{0+2}+\Sigma_k\Sigma_i V^k_i$).} 
\end{figure}

If the assumption of Gaussian likelihood holds, the $\chi^2$ values are expected to follow a $\chi^2$ distribution with degrees of freedom equal to the length of the data vector. In figure~\ref{fig:gaussian_test}, we plot the histogram (in blue) of the $\chi^2$ values measured from the Molino mock galaxy catalogs and compare it with the theoretical $\chi^2$ distribution curve for each of the summary statistics. Due to the existence of shot noise in the histogram, the curve may not agree with the histogram well, even for a sample strictly following the $\chi^2$ distribution. To visualize the effect of shot noise and help access the Gaussianity of the likelihood, we create 15000 multivariate Gaussian distributed data vectors with the same mean and covariance matrix as those estimated from the mock catalogs. We then obtain 15000 $\chi^2$ values for the multivariate Gaussian distributed data vectors and also plot a histogram (in red) for them in figure~\ref{fig:gaussian_test}.

As seen in the left panel of figure~\ref{fig:gaussian_test}, the histogram of $\chi^2$ values for the monopole and quadrupole of the power spectrum is very close to that for the Gaussian distributed data vectors and agree well with the theoretical $\chi^2$ distribution curve. This indicates that the likelihood of the power spectrum can be well modeled as Gaussian. For the MFs with four smoothing scales (the center panel) and the combination of the power spectrum and MFs (the right panel),  the agreement with the histogram of the random samples generated from the multivariate Gaussian and the theoretical curves is a little weaker. We find the non-Gaussianities mainly come from the MFs with $R_G=35 h^{-1}\rm{Mpc}$ and more non-Gaussianities exist in the likelihood for the MFs with larger smoothing scales, which is the reason why we only choose the smoothing scales satisfying $R_G<40 h^{-1}\rm{Mpc}$. The non-Gaussianities present in the MFs with large smoothing scales are expected: large $k$-modes are smeared out, and small $k$-modes are dominant in LSS when large smoothing scales are used; there are few small $k$-modes in a finite simulation box; thus the MFs of LSS dominated by a small number of large-scale $k$-modes are not expected to follow a Gaussian distribution. For upcoming redshift surveys whose survey volume is much larger than the simulation box, small $k$-modes are better sampled. Thus, the likelihood of the MFs with large smoothing scales is supposed to be well approximated by the Gaussian distribution. More extensive scales can be probed by the MFs without violating the Gaussian likelihood assumption.

\section{Convergence test}
\label{sec:conver}

\begin{figure}[tbp]
	\centering
	\includegraphics[width=1.0\textwidth]{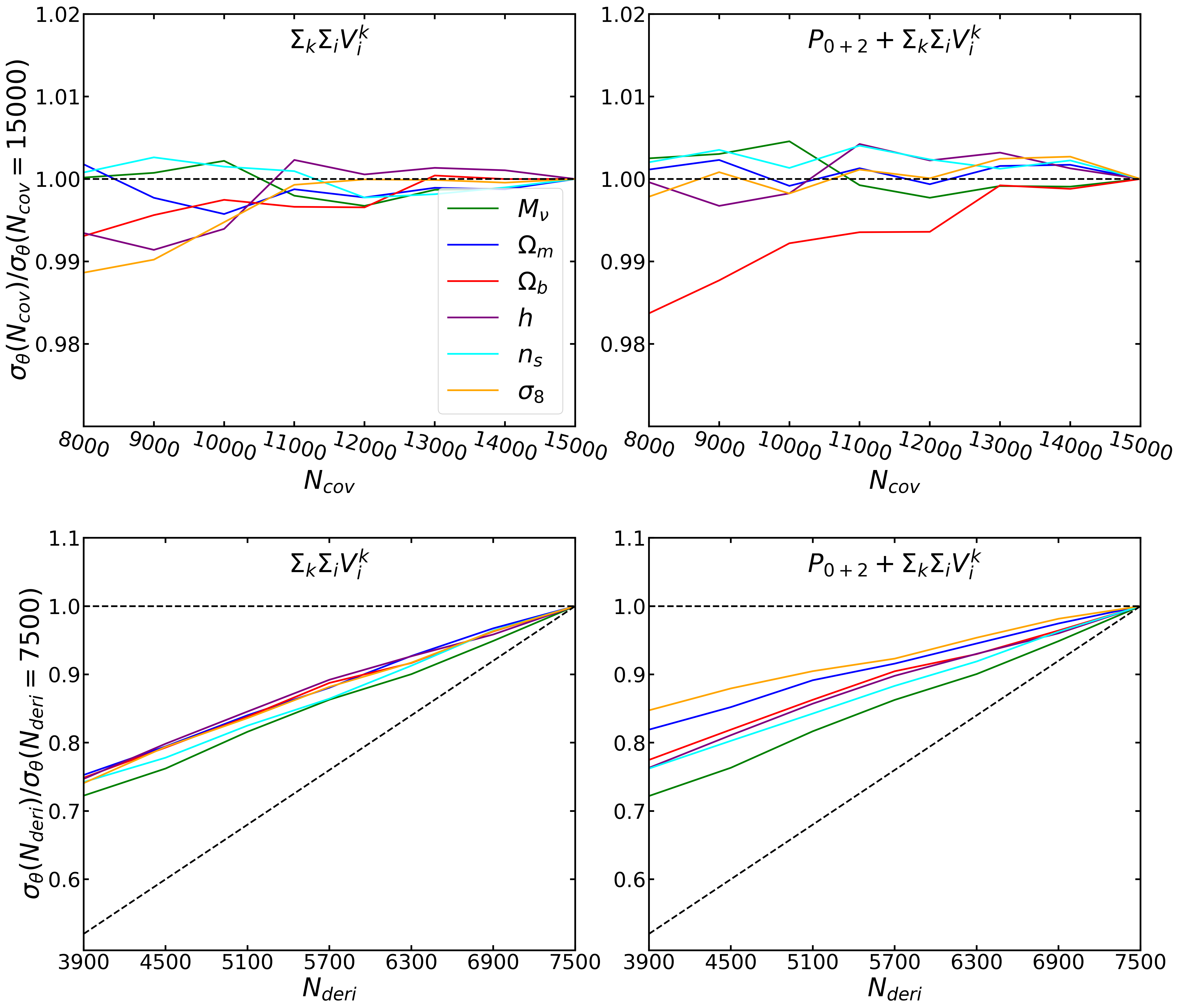}
	\caption{\label{fig:conver} Convergence of the marginalized errors from the combination of the MFs with four smoothing scales ($R_G=20,\ 25,\ 30,\ 35\ h^{-1}\mathrm{Mpc}$, $\Sigma_k\Sigma_i V^k_i$, the left column) and the combination of the power spectrum and the MFs ($P_{0+2}+\Sigma_k\Sigma_i V^k_i$, the right column) on $M_{\nu}$, $\Omega_{m}$, $\Omega_{b}$, $h$, $n_s$, and $\sigma_{8}$. 
The first row shows $\sigma_{\theta}(N_{cov})/\sigma_{\theta}(N_{cov}=15000)$, the ratio of Fisher forecasts obtained with covariance matrices estimated from $N_{cov}$ catalogs to those obtained with covariance matrices estimated from $N_{cov}=15000$ catalogs. Both forecasts are based on derivatives estimated with $N_{deriv}=7500$ catalogs. The horizontal dashed lines correspond to $\sigma_{\theta}(N_{cov})/\sigma_{\theta}(N_{cov}=15000) = 1$.
The second row shows $\sigma_{\theta}(N_{deri})/\sigma_{\theta}(N_{deri}=7500)$, the ratio of Fisher forecasts obtained with derivatives estimated from $N_{deri}$ catalogs to those obtained with derivatives estimated from $N_{deri}=7500$ catalogs. Both forecasts are based on covariance matrices estimated with $N_{cov}=15000$ catalogs. The horizontal dashed lines correspond to $\sigma_{\theta}(N_{deri})/\sigma_{\theta}(N_{deri}=7500) = 1$. And the oblique dashed line shows the ratio of the number of mocks used to the full sample $N_{deri}/7500$.} 
\end{figure}

When the cosmological parameter space has a high dimension, the degeneracies between parameters lead to a Fisher matrix with a large condition number. The parameter covariance matrix is obtained by inverting the Fisher matrix; thus, small uncertainties in the Fisher matrix may result in much larger deviations in the parameter covariance matrix. The uncertainties of the Fisher matrix stem from the noises existing in both the estimated covariance matrix and derivatives. In Figure~\ref{fig:conver}, we check how the marginalized errors vary when the number of catalogs used to estimate the covariance matrix $N_{\rm cov}$ or the derivatives $N_{\rm deri}$ is increased. We find $\sigma_{\theta}(N_{cov})$ converges very fast. When $N_{\rm cov}>8000$, it varies $\lesssim 2\%$ for all parameters, for both the constraints from the combination of the MFs with the four smoothing scales ($\Sigma_k \Sigma_i V^k_i$) and those from the combination of the power spectrum and the MFs ($P_{0+2}+\Sigma_k \Sigma_i V^k_i$). The convergence of $\sigma_{\theta}(N_{deri})$ is slower than $\sigma_{\theta}(N_{cov})$ when the number of samples is increased. For both the constraints from $\Sigma_k \Sigma_i V^k_i$ and $P_{0+2}+\Sigma_k \Sigma_i V^k_i$, $\sigma_{\theta}(N_{deri})$ varies $\lesssim 10\%$ for $N_{\rm deriv}>6000$, which is well converged. Better convergence can be achieved by generating more HOD realizations for each simulation used to calculate the derivatives w.r.t. the 11 parameters (6 cosmological parameters plus 5 HOD parameters). However, it will be computationally too expensive for this work because several times more samples may be needed to obtain the percent-level convergence. 

In \cite{2022arXiv220601619C}, two alternative methods are proposed to reduce the noises present in the derivatives and obtain more convergent results. One is to smooth the derivative with the Gaussian process (GP) \cite{10.7551/mitpress/3206.001.0001}, and another is to compress the summary statistics using the MOPED algorithm \cite{2018MNRAS.476L..60A} and then calculate the parameter forecast with the compressed observables. However, the Gaussian process smoothing can be so aggressive that important physical signatures may be smeared out \cite{2022arXiv220601619C}. In addition, the GP smoothing may artificially reduce the parameter constraints for statistics with a complex non-diagonal covariance matrix \cite{2022arXiv221012743H}. It may be the same case for the MFs since the covariance of the MFs has a complex structure as well. At the same time, the MOPED compression can be suboptimal so that the following Fisher forecast constraint will be larger than the truth \cite{2018MNRAS.476L..60A}. We leave a more detailed investigation of improving the convergence of the Fisher forecast using the MFs in a future study.

\section{Convergence test on $M_{\nu}$ derivatives}
\label{sec:derivative_test}
\begin{figure}[tbp]
	\centering
	\includegraphics[width=1.0\textwidth]{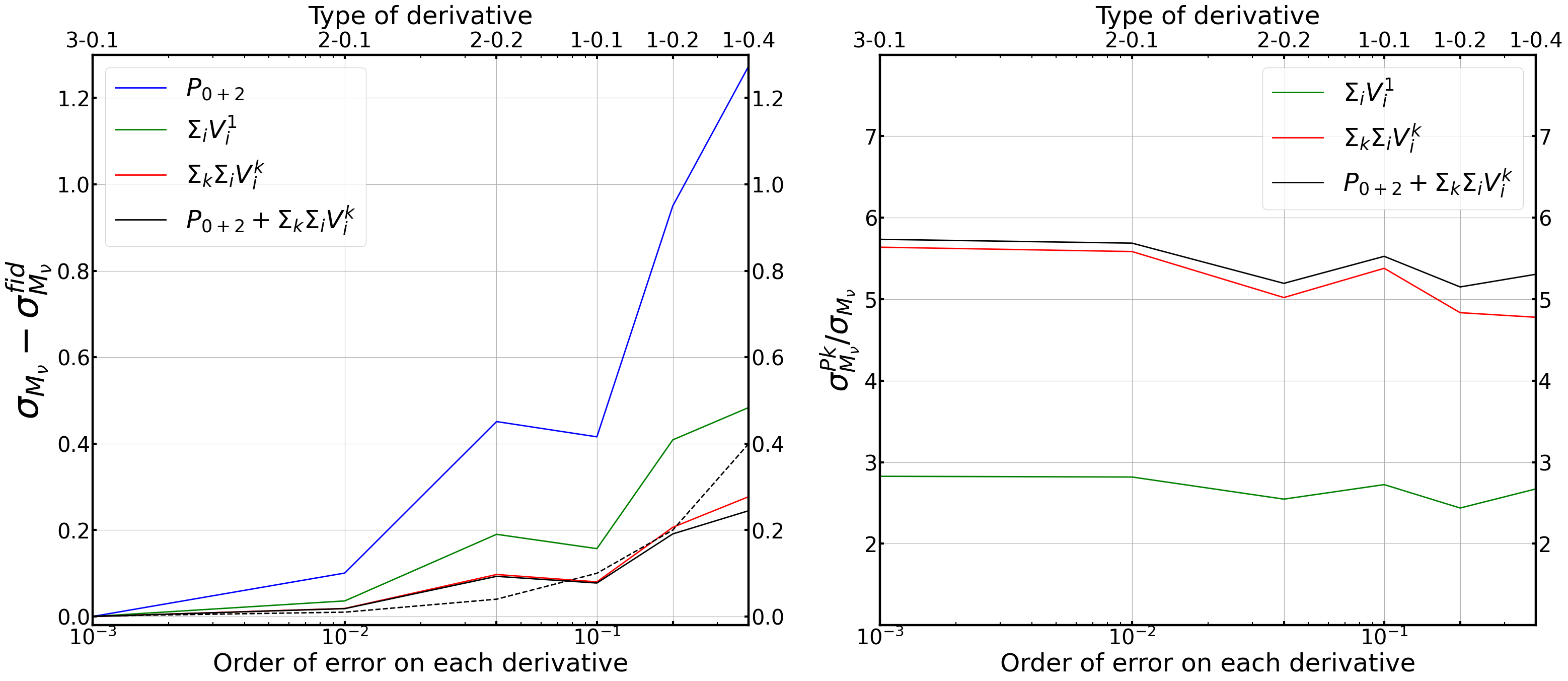}
	\caption{\label{fig:deri_est}Left panel: The difference in the constraint on $M_{\nu}$ (denoted as $\sigma_{M_{\nu}}$) and the fiducial constraint on $M_{\nu}$ (denoted as $\sigma^{fid}_{M_{\nu}}$, obtained with the derivative estimator $3-0.1$, the meaning of which will be explained below) as a function of the expected order of error for the derivative estimator w.r.t. $M_{\nu}$, or the type of the derivative estimator as shown in the top axis. These tick labels are in the format of $n-dM_{\nu}$, where $n$ is the index while $dM_{\nu}$ is the base of the big oh notations in equation~\ref{eq:nu_deri_1}, \ref{eq:nu_deri_2}, and~\ref{eq:nu_deri_3}. $\sigma_{M_{\nu}}-\sigma^{fid}_{M_{\nu}}$ is plotted for the monopole and quadrupole of the power spectrum ($P_{0+2}$, blue line), the MFs with $R_G=20h^{-1}\rm{Mpc}$ ($\Sigma_iV_i^1$, green line), the combination of the MFs with the fourth smoothing scales ($\Sigma_k \Sigma_i V_i^k$, red line), and the combination of the power spectrum and the MFs ($P_{0+2}+\Sigma_k \Sigma_i V_i^k$, black line). The black dashed line displays the expected order of magnitude for the error present in each derivative estimator. Right panel: The improvement of the power spectrum constraint on $M_{\nu}$ (denoted by $\sigma^{Pk}_{M_{\nu}}$) from the MFs with $R_G=20h^{-1}\rm{Mpc}$ ($\Sigma_iV_i^1$, green line), the combination of the MFs with the four smoothing scales ($\Sigma_k \Sigma_i V_i^k$, red line), and the combination of the power spectrum and the MFs ($P_{0+2}+\Sigma_k \Sigma_i V_i^k$, black line). The bottom and top x-axes are the same as the left panel. We note the improvement of $\sigma^{Pk}_{M_{\nu}}$ from the MFs and their combination with the power spectrum remains almost unchanged when estimators with varied accuracies are used.} 
\end{figure}

The derivative of observables w.r.t. $M_{\nu}$ can be estimated with the three kinds of estimators shown in equation~\ref{eq:nu_deri_1}, \ref{eq:nu_deri_2}, and \ref{eq:nu_deri_3}. We choose the third estimator~\ref{eq:nu_deri_3} in our forecast because it is the most accurate one. In figure~\ref{fig:deri_est}, we show how the constraint on $M_{\nu}$, $\sigma_{M_{\nu}}$, varies when types of derivative estimator change. The left panel shows the difference in $\sigma_{M_{\nu}}$ obtained with different types of derivatives and $\sigma^{fid}_{M_{\nu}}$ obtained with the most accurate derivative estimator, for the power spectrum ($P_{0+2}$), the MFs at $R_G=20h^{-1}\rm{Mpc}$ ($\Sigma_i V_i^1$), the combination of the MFs with four smoothing scales ($\Sigma_k\Sigma_i V_i^k$), and the combination of the power spectrum and the MFs ($P_{0+2}+\Sigma_k\Sigma_i V_i^k$). The expected order of magnitude for the error existing in each derivative estimator is plotted as a black dashed line. Although the propagation of uncertainties from the estimated derivatives to $\sigma_{M_{\nu}}$ is not straightforward, we know the error in $\sigma_{M_{\nu}}$ is proportional to the error in the $M_{\nu}$ derivative, hence the black dashed line may serve as an indicative reference. For the power spectrum, the actual error in $\sigma_{M_{\nu}}$ is a few times larger than that indicated by the dashed line. While for the MFs with one single smoothing scale or four varied smoothing scales, and their combination with the power spectrum, the actual error in $\sigma_{M_{\nu}}$ is quite close to the expected error.

In the right panel of figure~\ref{fig:deri_est}, we display the ratio of the power spectrum constraint on $M_{\nu}$ (denoted as $\sigma^{Pk}_{M_{\nu}}$) to $\sigma_{M_{\nu}}$ obtained from the MFs with one or four varied smoothing scales or the combination of the power spectrum and the MFs. The improvement of $\sigma^{Pk}_{M_{\nu}}$ from these observables remains almost unchanged when varied types of derivative estimators are used, which demonstrates our finding that tighter constraints can be obtained with the MFs is robust against systematics existing in different types of the derivative estimator. Specifically, the combination of the power spectrum and the MFs improve $\sigma^{Pk}_{M_{\nu}}$ by a factor of larger than five; $\sigma_{M_{\nu}}$ from the MFs with four smoothing scales is approximately five times tighter than $\sigma^{Pk}_{M_{\nu}}$; when only one single smoothing scale is used, the improvement on $\sigma^{Pk}_{M_{\nu}}$ from the MFs is better than two times for all types of derivative. 

\bibliographystyle{JHEP}
\bibliography{reference}
\end{document}